\documentclass[a4paper,11pt]{article}
\usepackage[usenames]{color}
\usepackage[normalem]{ulem}
\usepackage{amsfonts,amssymb}
\usepackage{theorem}
\usepackage{cancel}
\usepackage{pifont}
\usepackage[dvips]{graphicx}

\def\g5{\gamma_{_\chi}}

\def\red{\color{red}}
\def\black{\color{black}}
\newcommand{\evr}[1]{\red{#1~}\black}%
\newcommand{\comment}[1]{}

\begin{document}
\comment
{\sl
There is no such thing as an unsolvable  problem. \\
Either it is well-defined, \\ then the solution is
straightforward. \\
Or it is ill-posed, then it is not a problem. \\
(Free quote from Nicola Pintacuda)
\par\noindent
{\bf In this version the subtraction procedure is
simplified. General proofs are provided for cyclicity}
{
\par
\rightline{MIT, Nov. 20,  2016 }
\rightline{Revised Pisa, November 17,  2017 }
\par
\begin{center}
\evr{MASTER}
\end{center}
}    
}
\bf
\begin{center}{
 {\huge $\gamma_5$} in  Dimensional Regularization:
a Novel Approach
}
\footnote{\tt This work is supported in part by funds provided 
by the U.S. Department
of Energy (D.O.E.) under cooperative research agreement \#DE FG02-05ER41360}
\end{center}
\normalsize \rm
\large
\rm
\centerline{
Ruggero~Ferrari\footnote{e-mail: {\tt ruggferr@mit.edu}}$^{ab}$}
\normalsize
\smallskip
\begin{center}
$^a$
Center for Theoretical Physics\\
Laboratory for Nuclear Science
and Department of Physics\\
Massachusetts Institute of Technology\\
Cambridge, Massachusetts 02139\\
$^b$
INFN, Sezione di Milano\\
via Celoria 16, I-20133 Milano, Italy

(MIT-CTP4782, IFUM-1047-FT, May 2016)
\end{center}
\normalsize
\bf
\comment{
\evr{
\centerline{Comment on May 2, 2016}
\begin{quotation}
\huge
In this paper we discuss the case $D$ integer and
positive (Part I).
\par
In Part II we introduce the {\sl Generalized} Trace.
The analytic continuation in $D$ is done on the 
GT.
\end{quotation}
} 
}
\par
\normalsize
\rm
\begin{quotation}{\bf Abstract: }
A new Dimensional Regularization of  $\gamma_5$
is proposed. Cyclicity and Lorentz covariance are
enforced. The extension to generic dimension is based
on the integral representation of the trace of
gamma's, presented in a previous paper.
\end{quotation}
\newpage
\section{Introduction}
\label{sec:intro}
In a previous paper \cite{Ferrari:2014jqa} we have
introduced an integral representation in $D$ dimensions
of the  trace of $N$ gamma's and one $\gamma_\chi$ ($=\gamma_5$
for integer $D=4$ )
\begin{eqnarray}&&
Tr(\not\! p_{1} \not\! p_2\dots \not\! p_{N-1} \not\! p_{N} )
=
\int d^{^N} \bar c\, 
\exp \Big( 
\sum_{i<j=1}^N\bar c_i( p_{i}, p_{j}) \bar c_j\Big)
\label{intro.001a}
\\&& 
Tr(\not\! p_{1} \not\! p_2\dots \not\! p_{N-1} \not\! p_{N}\gamma_\chi )
=
i^{\frac{D(D-1)}{2}}  \int d^{^D}\xi\, d^{^N} \bar c\, 
\exp \Big(  \sum_{i=1}^{N}\bar c_i  (p_i)_\mu\xi_\mu 
\nonumber\\&& 
+\sum_{i<j=1}^N\bar c_i( p_{i}, p_{j}) \bar c_j\Big),
\label{intro.001b}
\end{eqnarray}
where $\mu=1\cdots D$  and $i,j=1\cdots N$ and $(p_i, p_j)= 
\sum_{\mu=1}^D(p_i)_\mu( p_j)_\mu $. $\bar c_i$ and
$\xi_\mu$ are {\sl real} Grassmannian variables \cite{berezin}; i.e.
\begin{eqnarray}
\int d\xi_\mu = 0, \qquad
\int d\xi_\mu~ \xi_\nu= \delta_{\mu\nu},\quad
\int d\bar c_i = 0, \qquad
\int d\bar c_i ~\bar c_j = \delta_{ij}
.
\label{intro.002}
\end{eqnarray}
 $\{p_1,\cdots, p_N\}$ are generic vectors (e.g.
momenta and polarization vectors) in $D$ dimensions. Finally
the normalization factor is chosen  to be
\begin{eqnarray}
\gamma_\chi^\dagger = \gamma_\chi, \qquad
\gamma_\chi^2 = 1
\label{intro.003}
\end{eqnarray}
for integer $D$. More references on  $\gamma_5$ and
its use in Dimensional Renormalization can be found in 
\cite{Ferrari:2014jqa}.
\par
Simple examples (for integer $D$) of eqs. (\ref {intro.001a}) and
(\ref{intro.001b}) can be easily given
\begin{eqnarray}&&
Tr(\not\!p_1\not\!p_2)=
\int d\bar c_2\,d \bar c_1\exp \Big[\bar c_1( p_1,p_2)\bar c_2 
\Big]= ( p_1,p_2)
\nonumber\\&& 
\Longrightarrow Tr({\mathcal I}) =1
\label{intro.003.1}
\end{eqnarray}
and for $D=4$
\begin{eqnarray}&&
Tr(\not\!p_1\not\!p_2\not\!p_3\not\!p_4\gamma_\chi)=
i^{\frac{D(D-1)}{2}} 
\int d^D\xi\, d^D\bar c\exp \Big(\sum_{j=1}^Dc_j (p_j)_\mu \xi_\mu
\nonumber\\&& 
+\sum_{i<j=1}^4\bar c_i( p_{i}, p_{j}) \bar c_j\Big)\Big|_{D=4}
\nonumber\\&& 
 =
\int d^D\xi\,\Big((p_1)_\mu \xi_\mu(p_2)_\nu \xi_\nu
(p_3)_\rho \xi_\rho (p_4)_\sigma \xi_\sigma\Big) 
(-i)^\frac{D(D-1)}{2}\Big|_{D=4}
\label{intro.003.2}
\nonumber\\&& 
 =
-
\epsilon_{\mu\nu\rho\sigma}(p_1)_\mu(p_2)_\nu (p_3)_\rho(p_4)_\sigma.
\end{eqnarray}
See Ref. \cite{E.R.Caianiello:1952ww} for
some early work on the relation between   
the trace of Dirac matrices and the Pfaffian. 
For the Pfaffian written as an integral over Grassmannian
variables see Ref. \cite{Jaffe:1987bf}.
\par
The formulae in eqs.  (\ref{intro.001a}) and (\ref{intro.001b})
are  nice formal interpolations on different values of the space-time
dimensions $D$.
The very existence of this integral representation 
is a hint  to search for a consistent management of 
$\gamma_\chi$ in generic $D$.
\par
In the usual matrix representation to move from  $D=3$
to  $D=4$ one has to redefine $\gamma_5$ i.e. from 
$\gamma_5=-i\gamma_1 \gamma_2\gamma_3 $ to 
$\gamma_5=-\gamma_1 \gamma_2\gamma_3\gamma_4$, 
which are unique for the chosen space dimension. 
While in eq. 
(\ref{intro.001b}) the completely
antisymmetric tensor emerges from the  generic expression
of the integration over $d^D \xi$, by taking $D=3$ or $D=4$.
The integral representation looks as the perfect tool to tackle
such a problem of continuation in $D$.
\par
In this paper 
Part I (integer $D$) is devoted to the
 generalization of  eq. (\ref{intro.001b}) to the case of multiple
$\gamma_\chi$ factors. Lorentz covariance and cyclicity are 
protected in the procedure. Then we consider the mechanism of
{\sl pairing}, i.e. we integrate over the Grassmann variables
pertinent to a pair of $\gamma_\chi$. After the mechanism of
pairing has removed all the pairs, the trace contains zero 
or at most one $\gamma_\chi$.
Finally we bring the last $\gamma_\chi$ to the far right of the trace:
the canonical form.
\par
The procedure will be cast in a set of very simple rules. In particular
the algebra for  integer
values of $D$ is 
\par
\begin{eqnarray}
\gamma_\chi\gamma_\mu= -(-)^D \gamma_\mu\gamma_\chi, \quad D\in {\cal N},
\label{int.004}
\end{eqnarray}
i.e. the algebra of the standard matrix representation.
\par
Unfortunately this very simple result cannot be extended to
generic values of $D$. In fact the algebra  
\par
\begin{eqnarray}
\gamma_\chi\gamma_\mu= q \gamma_\mu\gamma_\chi 
\label{int.004.1 }
\end{eqnarray}
implies
\begin{eqnarray}
\gamma_\chi\gamma_\mu\gamma_\mu = q^2 \gamma_\mu \gamma_\mu\gamma_\chi ,
\label{int.004.2 }
\end{eqnarray}
i.e. 
\begin{eqnarray}
 q^2 =1,
\label{int.004.3 }
\end{eqnarray}
which forbids to continue eq. (\ref{int.004}) to complex $D$.
\par
This negative result is mitigated
in some  explicit calculations by the fact that  it is not necessary to know
the explicit form of 
\begin{eqnarray}
\Big\{\gamma_\chi,\gamma_\mu\Big\};
\label{int.004.4}
\end{eqnarray}
instead one can use 
\begin{eqnarray}
Tr\Big(\Big\{\gamma_\chi,\gamma_\mu\Big\}\gamma_{\mu_1}
\dots \gamma_{\mu_k}\Big)
=
Tr\Big(\gamma_\chi\Big\{\gamma_\mu,\gamma_{\mu_1}
\dots \gamma_{\mu_k}\Big\}\Big).
\label{int.005}
\end{eqnarray}
\par
On the basis of this assumption (i.e. the existence of an expansion
in powers of $(D-4)$) we have  tested
Dimensional Regularization  by explicit calculations:
 i) in \cite{Ferrari:2014jqa}
the ABJ anomaly 
\cite{Adler:1969gk} \cite{Bell:1969ts} and  the invariance of the path integral
functional (Local Functional Equation 
\cite{Ferrari:2005ii}-\cite{Bettinelli:2007kc} );
ii) in  \cite{Ferrari:2015mha} the isoscalar anomaly in a non abelian $SU(2)$ gauge
theory  
\cite{Bardeen:1969md}   \cite{bertlmann};
iii) in \cite{Ferrari:2015xba} the Chern-Simons term and photon self-energy in
QED with a CPT- and Lorentz-violating action term
\cite{Jackiw:1999yp} \cite{Jackiw:1999qq}.
\par
In all the listed calculations the technique has been 
very successful: no ambiguity emerges and the results
coincide with those present in the literature, 
obtained via gauge invariant regulators
(Pauli-Villars). 
\par
However in all the cases mentioned above the Feynman amplitude
is finite for ($D=4$), i.e. the pole in $D=4$ is cancelled 
by the zero emerging from  the $\gamma$'s algebraic manipulations
based on 
\begin{eqnarray}
\Big\{\gamma_\mu,\gamma_\nu\Big\} = 2\delta_{\mu\nu}.
\label{int.004p}
\end{eqnarray}
\par
It would be very interesting to extend the representation
of $\gamma_5$  to generic $D$ in any situation. For instance
in the case of a single (divergent) graph it is very
helpful to have a consistent Dimensional Regularization
method. This would allow formal algebraic manipulations.
A similar situation  occurs in the calculation of
a divergent Feynman amplitude, where a renormalization 
procedure is required, possibly via pole subtraction.
The aim of the
paper is to provide this tool.
\par
In Part II (non-integer $D$)
we suggest a novel procedure of Dimensional Renormalization: the pole removal 
must be performed 
\underline{before} the final integration over $\xi$ (any generic Grassmannian
variable that generates $\gamma_\chi$ when integrated over). 
In this way we get rid of the completely antisymmetric
tensor in the subtraction procedure and of the problem of its
analytic continuation. Moreover the algebra in eq. (\ref{int.004})
is implemented when the limit of integer $D$ is taken.
The mechanism of pairing can also be implemented with some 
simple modifications.
Thus the procedure is straightforward
down to the final step, when the completely antisymmetric tensor is
recovered by the integration over $\xi$.
\par
In conclusion of the paper  the
rules will be tested on  the example developed in  \cite{Ferrari:2014jqa}
for ABJ anomaly.
\section{Standard Identities for Integer $D$}
\label{sec:origin}
We recollect some standard  properties of $\gamma_5$.
\par\noindent
\underline{Let $D$ be integer and odd.}
Thus we can consider the product
\begin{eqnarray}
\gamma_5\equiv \prod_{\mu=1}^D \gamma_\mu.
\label{origin.1}
\end{eqnarray}
By using the gamma's algebra
\begin{eqnarray}
\{\gamma_\mu, \gamma_\nu\}=2\delta_{\mu\nu} , \qquad \forall \mu,\nu=1\dots D
\label{origin.1.1}
\end{eqnarray}
one gets
\begin{eqnarray}
[\gamma_5, \gamma_\mu]=0 , \qquad \forall \mu=1\dots D.
\label{origin.2}
\end{eqnarray}
From Schur's lemma we get that $\gamma_5$ is a number.
\par\noindent
\underline{Let $D$ be integer and even.}
Thus we can consider the product
\begin{eqnarray}
\gamma_5\equiv \prod_{\mu=1}^D \gamma_\mu
\label{origin.4}
\end{eqnarray}
and we get
\begin{eqnarray}
\{\gamma_5, \gamma_\mu\}=0 , \qquad \forall \mu=1\dots D.
\label{origin.5}
\end{eqnarray}
\par
Now we recall some further  identities obtained from eq. (\ref{origin.1.1})
\begin{eqnarray}&&
\gamma_{\mu}\gamma_\rho\gamma_{\mu}=(-D+2)\gamma_\rho
\nonumber\\&&
\gamma_{\mu}\gamma_\rho\gamma_\sigma\gamma_{\mu}=(D-4)\gamma_\rho\gamma_\sigma+ 4\delta_{\rho\sigma}
\nonumber\\&&
\gamma_{\mu}\gamma_{\mu_1}\dots \gamma_{\mu_{k}}\gamma_{\mu}
=(-)^k
(D-2k)\gamma_{\mu_1}\dots \gamma_{\mu_{k}}
\nonumber\\&&
+4(-)^k\sum_{\{ij\}}\delta_{\cal P} \delta_{\mu_i \mu_j}\gamma_{\mu_1}\dots{\hat\gamma_{\mu_i}}
\dots{\hat\gamma_{\mu_j}}\dots \gamma_{\mu_{k}},
\label{int.8p}
\end{eqnarray}
where $~\hat{}$ means omitted and $\delta_{\cal P}$ the parity of
the permutations to take them in front.

\part{ Integer $D$}
\label{part.1}
\section{The Use of Cyclicity}
\label{sec:cycle}
When $\gamma_\chi$ is absent, as in eq. (\ref{intro.001a}),
cyclicity is a property of the trace. The proof is
given in Ref. \cite{Ferrari:2014jqa}. 
\par
If $\gamma_\chi$ is introduced as in  eq. (\ref{intro.001b})
then cyclicity can not be deduced, since its very definition 
is fixed by its position. Instead, if we \underline{
require} this property, then we can extend the expression
in   eq. (\ref{intro.001b}) to the cases, where the position
of $\gamma_\chi$ is generic in the trace expression.
\par
To illustrate this fact let us consider the identity
\begin{eqnarray}&&
i^{\frac{D(D-1)}{2}}  \int d^{^D}\chi\, d^{^N} \bar c\, \,
2[\bar c_k (p_k, p_N)\bar c_N]
\exp \Big(  \sum_{i=1}^{N}\bar c_i  (p_i)_\mu\chi_\mu 
\nonumber\\&& 
+\sum_{i<j=1}^N\bar c_i( p_{i}, p_{j}) \bar c_j\Big)
= \delta_P
Tr(\not\! p_{1}\dots \big\{ \not\! p_k ,\not\! p_{N}\big\}\dots  \not\! p_{N-1}\gamma_\chi )
,
\label{cycle.1}
\end{eqnarray}
where $\delta_P= (-)^{N-k-1}$ is the parity of the permutations necessary to order 
the factors in the form: 
\begin{eqnarray}
d\bar c_N \bar c_N \,\,\, d \bar c_k \bar c_k.
\label{cycle.2}
\end{eqnarray}
Now we use the identity
\begin{eqnarray}&&
\not\! p_{N} \not\! p_{1}\dots   \not\! p_{N-1}
= \sum_{k=1}^{N-1}\not\! p_{1}\dots \big\{ \not\! p_k ,\not\! p_{N}\big\}\dots  \not\! p_{N-1} (-)^{k-1}
\nonumber\\&& 
+(-)^{N-1 }\not\! p_{1}\dots   \not\! p_{N-1}\not\! p_{N}
\label{cycle.3.0}
\end{eqnarray}
and  sum over $k$ in eq. (\ref{cycle.1})
\begin{eqnarray}&&
i^{\frac{D(D-1)}{2}}  \int d^{^D}\chi\, d^{^N} \bar c\, \,
2\sum_{k=1}^{N-1}[\bar c_k (p_k, p_N)\bar c_N]
\exp \Big(  \sum_{i=1}^{N}\bar c_i  (p_i)_\mu\chi_\mu 
\nonumber\\&& 
+\sum_{i<j=1}^N\bar c_i( p_{i}, p_{j}) \bar c_j\Big)
= 
(-)^NTr\Big(\not\! p_{N}\not\! p_{1}\dots   \not\! p_{N-1}\gamma_\chi \Big)
\nonumber\\&& 
+ Tr\Big (  \not\! p_{1}\dots   \not\! p_{N-1}\not\! p_{N}\gamma_\chi \Big).
\label{cycle.3}
\end{eqnarray}
By using cyclicity one gets
\begin{eqnarray}&&
i^{\frac{D(D-1)}{2}}  \int d^{^D}\chi\, d^{^N} \bar c\, \,
2\sum_{j=1}^{N-1}[\bar c_j (p_j, p_N)\bar c_N]
\exp \Big(  \sum_{i=1}^{N}\bar c_i  (p_i)_\mu\chi_\mu 
\nonumber\\&& 
+\sum_{i<j=1}^N\bar c_i( p_{i}, p_{j}) \bar c_j\Big)
=
Tr\Big(\not\! p_{1}\dots   \not\! p_{N-1}\not\! p_{N}\gamma_\chi \Big)
\nonumber\\&& 
+ 
 (-)^N Tr\Big(\not\! p_{1}\dots   \not\! p_{N-1}\gamma_\chi \not\! p_{N}\Big).
\label{cycle.4}
\end{eqnarray}
Eq. (\ref{cycle.4}) suggests how to represent the trace
when $\gamma_\chi$ is in second position.
\par
To illustrate this we consider
\begin{eqnarray}&&
Tr\Big(\not\! p_{1}\dots   \not\! p_{N-1}\gamma_\chi \not\! p_{N}\Big)
=
i^{\frac{D(D-1)}{2}}  \int d\bar c_N d^{^D}\chi\, d^{^{(N-1)}} \bar c\, \,
\nonumber\\&& 
\exp \Big(\chi_\mu p_{_N\mu}\bar c_N + \sum_{i=1}^{N-1}\bar c_i  (p_i)_\mu\chi_\mu 
+\sum_{i<j=1}^N\bar c_i( p_{i}, p_{j}) \bar c_j\Big)
\label{cycle.4.1}
\end{eqnarray}
Perform the switch $ d\bar c_N d^{^D}\chi\to (-)^{_D} 
d^{^D}\chi d\bar c_N $ and the change of sign $\chi \to - \chi$
\begin{eqnarray}&&
= i^{\frac{D(D-1)}{2}} \int  d^{^D}\chi\,d\bar c_N d^{^{(N-1)}} \bar c\, \,
\exp \Big( -\sum_{i=1}^{_{N-1}}\bar c_i  p_i\chi
+ p_{_N\mu}\bar c_N \chi_\mu
\nonumber\\&& 
+\sum_{i<j=1}^{N-1}\bar c_i( p_{i}, p_{j}) \bar c_j
+\sum_{j=1}^{N-1}\bar c_j( p_{j}, p_{N}) \bar c_N \Big)
\label{cycle.4.2}
\end{eqnarray}
Change sign $ d\bar c_j \to - d\bar c_j ~~\forall j<N$
\begin{eqnarray}&&
=i^{\frac{D(D-1)}{2}} (-)^{_{N-1}}  \int d^{^D}\chi\, d\bar c_N  d^{^{(N-1)}} \bar c\, \,
\exp \Big( \sum_{i=1}^{_{N}}\bar c_i  (p_i)\chi 
\nonumber\\&& 
+\sum_{i<j=1}^{_{N-1}}\bar c_i( p_{i}, p_{j}) \bar c_j
-\sum_{j=1}^{_{N-1}}\bar c_j( p_{j}, p_{N}) \bar c_N \Big)
\label{cycle.4.3}
\end{eqnarray}
Make a shift $ \bar c_{_{N}}\to \bar c_{_{1}},  \bar c_j \to \bar c_{j+1} 
 ~~\forall j<N$
\begin{eqnarray}&&
=i^{\frac{D(D-1)}{2}}  \int d^{^D}\chi\,d\bar c_N \, d^{^{(N-1)}} \bar c\, \,\exp 
\Big(\sum_{i=1}^{_{N}}\bar c_i  (p_i)\chi 
\nonumber\\&& 
+\sum_{i<j=1}^{_{N-1}}\bar c_{_{i+1}}( p_{i}, p_{j}) \bar c_{_{j+1}}
-\sum_{j=1}^{_{N-1}}\bar c_{_{j+1}}( p_{j}, p_{N}) \bar c_1
\Big)
\nonumber\\&& 
=i^{\frac{D(D-1)}{2}}  \int d^{^D}\chi\,d\bar c_N \, d^{^{(N-1)}} \bar c\, \,\exp 
\Big(\sum_{i=1}^{_{N}}\bar c_i  (p_i)\chi 
\nonumber\\&& 
+\sum_{i<j=2}^{_{N}}\bar c_{_{i}}( p_{_{i-1}}, p_{{_{j-1}}} )\bar c_{_{j}}
+\bar c_1\sum_{j=2}^{_{N}}\bar c_{_{j}}( p_{_{j-1}}, p_{N}) 
\Big)
\nonumber\\&& 
=
Tr\Big(\not\! p_{_N}\not\! p_{1}\dots   \not\! p_{_{N-1}}\gamma_\chi \Big).
\label{cycle.5}
\end{eqnarray}
cyclicity works in this example! 
\section{More $\gamma_\chi$'s}
\label{sec:more}
The generalization of eq. (\ref{intro.001b}) to more than one 
$\gamma_\chi$ is achieved by following the suggestion of 
the result in eq. (\ref{cycle.5}) and by the method of unfolding $\gamma_\chi$
into a product of $D$ all different $\gamma$'s for integer dimensions:
\begin{eqnarray}
\gamma_\chi = (i)^\frac{(D-1)D}{2}\gamma_1\cdots \gamma_D.
\label{more.1}
\end{eqnarray}
Thus we can write the trace as it were  with no $\gamma_\chi$.
See Section 6 in Ref.  \cite{Ferrari:2014jqa} for details.
\par\noindent
{\bf Assumption 1:} Multiple $\gamma_\chi$ trace is represented
by integration over Grassmannian variables. Each $\not\!\! p_j$
is associated to the integration variable  $d \bar c_j$
while $\gamma_\chi$'s are represented by integration
over $\dots d^D \chi \dots d^D \eta
\dots d^D \xi \dots$. The order in the trace is
faithfully reproduced in the order of integration and in the
terms entering in the exponential. A factor 
\begin{eqnarray}
 (i)^\frac{(D-1)D}{2}
\label{more.1.1}
\end{eqnarray}
is introduced for each $\gamma_\chi$.
\section{Example with Two $\gamma_\chi$}
\label{sec:two}
We illustrate the algorithm in  the case of two $\gamma_\chi$.
\begin{eqnarray}&&
Tr(\not\! p_{1} \dots \not\! p_k\gamma_\chi \cdots\not\! p_{N} \gamma_\chi )
=
(-)^{\frac{D(D-1)}{2}}  \int d^{^D}\eta\, d^{^{(N-k)}} \bar c\, d^{^D}\xi d^k \bar c
\nonumber\\&&
\exp \Big(  \sum_{i=1}^{N}\bar c_i  (p_i)_\mu\eta_\mu +\xi_\mu\eta_\mu
+\xi_\mu \sum_{i=k+1}^{N}\bar c_i  (p_i)_\mu +\sum_{i=1}^{k}\bar c_i  (p_i)_\mu\xi_\mu
\nonumber\\&&
+\sum_{i<j=1}^N\bar c_i( p_{i}, p_{j}) \bar c_j\Big)
\nonumber\\&&
=
(-)^{\frac{D(D-1)}{2}}  \int d^{^D}\eta\, d^{^{(N-k)}} \bar c\, d^{^D}\xi d^k \bar c
\nonumber\\&&
\exp \Big(  \sum_{i=1}^{N}\bar c_i  (p_i)_\mu\eta_\mu
+\xi_\mu\Big[\eta_\mu+ \sum_{i=k+1}^{N}\bar c_i  (p_i)_\mu -\sum_{i=1}^{k}\bar c_i  (p_i)_\mu\Big]
\nonumber\\&&
+\sum_{i<j=1}^N\bar c_i( p_{i}, p_{j}) \bar c_j\Big).
\label{more.2.1}
\end{eqnarray}
Now we perform the integration over both $\xi,\eta$. The procedure is
to force the integration over $\xi,\eta$ by an isolated term $\xi\eta$
in the exponential. Only the $\xi_\mu$ in $\xi\eta$ can saturate the integration
over $\xi_\mu$ which has a factor $\eta_\mu$. Therefore also the integration
over $\eta$ is constrained
\begin{eqnarray}&&
\int d^D\eta d^D\xi \, e^{\xi\eta} = \int d^D\eta d^D\xi \, \prod_\mu(1+\xi_\mu\eta_\mu)
\nonumber\\&&
=  \int d^D\eta d^D\xi \, \prod_\mu\xi_\mu\eta_\mu
=(-)^{\frac{D(D-1)}{2}}.
\label{more.2.2}
\end{eqnarray}
In order to employ the procedure of eq. (\ref{more.2.2}) we replace in 
eq. (\ref{more.2.1})
\begin{eqnarray}
\eta \to \eta -  \sum_{i=k+1}^{N}\bar c_i  (p_i) +\sum_{i=1}^{k}\bar c_i  (p_i)
\label{more.2.3}
\end{eqnarray}
and we get
\begin{eqnarray}&&
Tr(\not\! p_{1} \dots \not\! p_k\gamma_\chi \cdots\not\! p_{N} \gamma_\chi )
=
(-)^{\frac{D(D-1)}{2}}  \int d^{^D}\eta\, d^{^{(N-k)}} \bar c\, d^{^D}\xi d^k \bar c
\nonumber\\&& 
\exp \Big(  \sum_{i=1}^{N}\bar c_i  (p_i)_\mu\eta_\mu 
-2[\sum_{i=1}^{k}\bar c_i  (p_i)]  [\sum_{i=k+1}^{N}\bar c_i  (p_i)]
\nonumber\\&& 
+\xi\eta
+\sum_{i<j=1}^N\bar c_i( p_{i}, p_{j}) \bar c_j\Big),
\label{more.2.4}
\end{eqnarray}
Now the integration over $\xi\eta$ can be performed
\begin{eqnarray}&&
Tr(\not\! p_{1} \dots \not\! p_k\gamma_\chi \cdots\not\! p_{N} \gamma_\chi )
=
(-)^{\frac{D(D-1)}{2}}  (-)^{(D(N-k))} (-)^{\frac{D(D-1)}{2}}
\nonumber\\&& 
\int  d^{^N} \bar c
\exp \Big( 
-[\sum_{i=1}^{k}\bar c_i  (p_i)]  [\sum_{i=k+1}^{N}\bar c_i  (p_i)]
\nonumber\\&&
+ \sum_{i<j=1}^k\bar c_i( p_{i}, p_{j}) \bar c_j
+\sum_{i<j=k+1}^N\bar c_i( p_{i}, p_{j}) \bar c_j\Big).
\label{more.2.5}
\end{eqnarray}
Finally we change sign to $\bar c_j, j=k+1,\dots, N$.
\begin{eqnarray}&&
Tr(\not\! p_{1} \dots \not\! p_k\gamma_\chi \cdots\not\! p_{N} \gamma_\chi )
=  (-)^{(D-1)(N-k)}
\nonumber\\&& 
\int  d^{^N} \bar c
\exp \Big( 
[\sum_{i=1}^{k}\bar c_i  (p_i)]  [\sum_{i=k+1}^{N}\bar c_i  (p_i)]
+ \sum_{i<j=1}^k\bar c_i( p_{i}, p_{j}) \bar c_j
+\sum_{i<j=k+1}^N\bar c_i( p_{i}, p_{j}) \bar c_j\Big)
\nonumber\\&&
=
(-)^{(D-1)(N-k)}
Tr(\not\! p_{1} \dots \not\! p_k \cdots\not\! p_{N} )
.
\label{more.2.6}
\end{eqnarray}
%
In particular for $k=N$ we get 
\begin{eqnarray}&&
Tr(\not\! p_{1} \dots \not\! p_{N} \gamma_\chi^2 )
=  
Tr(\not\! p_{1} \dots \not\! p_{N}  )
.
\label{more.2.7}
\end{eqnarray}
Notice that the result in eqs. (\ref{more.2.6}) and  (\ref{more.2.7})
coincides with
the standard  algebra of the gamma matrices (integer dimensions).
\par
This integration over a couple of variables describing 
a pair of $\gamma_\chi$ will be denoted as {\sl pairing}.
\par
We shall provide more examples involving three $\gamma_\chi$.
%
\section{Example with three $\gamma_\chi$}
\label{sec:three}
It is very instructive to consider a case with an odd
number of $\gamma_\chi$. After the use of the tool of
pairing, only one $\gamma_\chi$ is left over at the end. 
\par
We consider the explicit example where three $\gamma_\chi$
are present in the trace.
The order among the
Grassmannian variable follows faithfully the order inside
the trace as in eq. (\ref{more.2.1})
\begin{eqnarray}&&
Tr(\not\! p_{1} \not\! p_2\dots\gamma_\chi \not\! p_{_{N-2}}  \not\! p_{_{N-1}}\gamma_\chi  \not\! p_{_{N}}\gamma_\chi )
\nonumber\\&&  
=
(-i)^{\frac{D(D-1)}{2}}  \int d^{^D}\chi\,d\bar c_{_N}\, d^{^D}\xi\, d\bar c_{_{N-1}}\,
d\bar c_{_{N-2}}\,d^{^D}\eta\, d^{^{N-3}} \bar c\, 
\exp \Big\{\Big[\sum_{i=1}^{N}\bar c_i  (p_i) 
\nonumber\\&&  
  +\xi  + \eta \Big]\chi 
+\Big[\xi   +\eta \Big] \bar c_{_N}  (p_N)   +\Big[ \eta  
+\sum_{i=1}^{_{N-1}}\bar c_i ( p_i) \Big]\xi 
\nonumber\\&&  
+\eta  \Big(\bar c_{N-1}(p_{N-1})  + \bar c_{N-2}(p_{N-2}) \Big)
+ \sum_{i=1}^{_{N-3}}\bar c_i ( p_i) \eta  
\nonumber\\&&  
+\sum_{i=1,i<j\le N}^N\bar c_i( p_{i}, p_{j}) \bar c_j\Big\}.
\label{more.5.1}
\end{eqnarray}
The question 
arises on the relations among the different paths of pairings.
Thus we explore some possibilities. The calculation is 
straightforward and similar to the previous leading
to eq. (\ref{more.2.6}).
\subsection{Integration over  $\chi_\mu$ and $\eta_\mu$}
If we want to perform the pairing
by integration over $\chi_\mu$ and $\eta_\mu$ it is easier to isolate
the factor $\sum_\mu \eta_\mu\chi_\mu$ so that we get rid of all the integration
$d^D\chi \, d^D\eta$. This is achieved by the substitution
$\eta_\mu \to \eta_\mu -\xi_\mu -\sum_{i=1}^{N}\bar c_i  (p_i)_\mu$.
Then one can  show that
\begin{eqnarray}&&
Tr(\not\! p_{1} \not\! p_2\dots\gamma_\chi \not\! p_{_{N-2}}  \not\! p_{_{N-1}}\gamma_\chi  \not\! p_{_{N}}\gamma_\chi )
\nonumber\\&&  
=(-)^{3(D-1)}
Tr(\not\! p_{1} \not\! p_2\dots \not\! p_{_{N-2}}  \not\! p_{_{N-1}}\gamma_\chi  \not\! p_{_{N}} ).
\label{more.8pp}
\end{eqnarray}
%
\subsection{Integration over  $\chi_\mu$ and $\xi_\mu$}
For the integration over $d^D\chi d^D\xi $ we proceed as 
in the previous case: we isolate a term $\xi\chi$
by a substitution $\xi \to \xi-\eta -\sum_{i=1}^N\bar c_i p_i$.
Explicit calculation yields
\begin{eqnarray}&&
Tr(\not\! p_{1} \not\! p_2\dots\gamma_\chi  \not\! p_{_{N-2}} 
\not\! p_{_{N-1}}\gamma_\chi  \not\! p_{_{N}}\gamma_\chi )
\nonumber\\&&  
=(-)^{D-1}
Tr(\not\! p_{1} \not\! p_2\dots\gamma_\chi  \not\! p_{_{N-2}} 
\not\! p_{_{N-1}} \not\! p_{_{N}}).
\label{more.7.1p}
\end{eqnarray}
%
\subsection{Integration over  $\xi\, \eta$}
%
Finally we  consider the last pairing  in eq. (\ref{more.5.1})
by integrating over  $ d^D \xi\, d^D \eta$.
We isolate a common factor $\exp \eta\chi$ by using the substitution
$\eta_\mu \to \eta_\mu +\chi_\mu+\bar c_N(p_N)_\mu
- \sum_{i=1}^{N-1}\bar c_i (p_i)_\mu$.
One gets
\begin{eqnarray}&&
Tr(\not\! p_{1} \not\! p_2\dots\gamma_\chi  \not\! p_{_{N-2}}
\not\! p_{_{N-1}}\gamma_\chi  \not\! p_{_{N}}\gamma_\chi )
\nonumber\\&&  
=
 (-)^{2(D-1)} 
Tr(\not\! p_{1} \not\! p_2\dots  \not\! p_{_{N-2}}
\not\! p_{_{N-1}} \not\! p_{_{N}}\gamma_\chi )
\label{more.8.3p}
\end{eqnarray}
We see that,  in all the example
presented, pairing is obtained by using the naive algebra
\begin{eqnarray}&&
\gamma_\chi\gamma_\mu = (-)^{D-1} \gamma_\mu \gamma_\chi
\nonumber\\&&
\gamma_\chi ^2 = 1. 
\label{more.8.1.1}
\end{eqnarray}
The final result of the pairing process is independent
from the chosen sequence, as one can verify by using
eqs. (\ref{more.8pp}), (\ref{more.7.1p}),  (\ref{more.8.3p})
and (\ref{more.8.1.1}).
One can  prove  that the properties in eq. (\ref{more.8.1.1}) 
are true for any set of $\gamma$ and $\gamma_\chi$
for integer dimensions $D$. Moreover under the same
conditions one can prove cyclicity.

\section{Algebra and Cyclicity in General for Integer $D$}
\label{sec:morecycl}
%
One can  prove  that the properties in eq. (\ref{more.8.1.1}) 
are true for any set of $\gamma_\mu$ and $\gamma_\chi$
for integer dimensions $D$. Moreover under the same
conditions one can prove cyclicity 
\begin{eqnarray}
Tr\Big( {\not\!p_1\cal A}\Big)
=Tr\Big( {\cal A}\not\!p_1\Big) ,
\label{morecycl.1}
\end{eqnarray}
where $\cal A$ is any product of $\not p$ and $\gamma_\chi$.
\par
Similarly one has
\begin{eqnarray}
Tr\Big( {\gamma_\chi\cal A}\Big)
=Tr\Big( {\cal A}\gamma_\chi\Big).
\label{morecycl.2}
\end{eqnarray}
%
%
\newpage
\part{New Approach to {\huge$\gamma_5$} in DR}

In the Introduction it was argued that in the equations 
(\ref{more.8.1.1}) $D$ cannot be continued to complex values. 
Moreover the completely antisymmetric tensor is 
identified by the number of the dimensions and therefore it is a
quantity that cannot depend smoothly on $D$. These are  
insurmountable obstacles for any  continuation in
$D$.
\par
In this paper we suggest a way out to this {\sl impasse}.
We take full advantage of the integral representation
of the trace in eq. (\ref{intro.001b}) and its generalizations
with more than one $\gamma_\chi$ discussed in Part \ref{part.1}.
We review the properties of this representation that are not
dependent from $D$ being an integer. In particular we shall
avoid by any means the use of quantities as $(-1)^{_D}$ in 
the algebra, as in eq.(\ref{more.8.1.1}), because
they are an uncontrolled source of ambiguities.

\section{Pairing}
\label{sec:gener}
We consider the most general pairing setup. We evaluate
\begin{eqnarray}&&
Tr\Big( {\cal A}\gamma_\chi {\cal B} \gamma_\chi{\cal C}\Big)
=(i)^{(D-1)D}\int\, d{\cal C}\, d^D\xi\, d{\cal B}\, d^D \eta\,  d{\cal A}
\nonumber\\&&
\exp\Bigg({\cal C}*{\cal C} + \big(\xi+{\cal B}+\eta +{\cal A}  \big){\cal C}
\nonumber\\&&
+\big({\cal B}+\eta +{\cal A}  \big)\xi
+{\cal B}*{\cal B}+ \big(\eta +{\cal A} \big){\cal B}
+{\cal A} \eta + {\cal A}*{\cal A}
\Bigg)
\label{gener.2}
\end{eqnarray}
where ${\cal A}=\{a_1,\dots,a_K\}$ and ${\cal B}=\{b_1,\dots,b_L\}$
are sets of elements $\bar c_i(p_i)_\mu$; while  ${\cal C}=\{c_1,\dots,c_M\}$
may contain both $\bar c_i(p_i)_\mu$ and the Grassmannian integration
variables $\chi_\mu$. 
For generic ${\cal A}$ the traces and products are defined by
(example with one $\gamma_\chi$ only)
\begin{eqnarray}&&
Tr\big( {\cal A}\big) = Tr\Big(\not\!p_1\dots\not\!p_{n}\gamma_\chi
\not\!p_{{ n+1}}\dots\not\!p_K\Big).
\label{gener.3}
\end{eqnarray}
in the matrix representation. In the integral representation
we use a short hand notation which follows from the notations
in Sections \ref{sec:more}, \ref{sec:two} and \ref{sec:three}
\begin{eqnarray}&&
{\cal A}*{\cal A} = \sum_{i<j=1}^K \bar c_i(p_i, p_j) \bar c_j
+ \sum_{n<j}^K(\chi, p_j) \bar c_j + \sum_{j=1}^{n}  \bar c_j(p_j,\chi)
\nonumber\\&&
-{\cal A} = \{-a_1,\dots,-a_K\}\qquad
{\cal A} \chi = \sum_{i=1}^K \bar c_i (p_i,\chi).
\label{gener.3.1}
\end{eqnarray}
The differentials $d{\cal A}$ are defined by the product of
$d\bar c_i~ (i=1\dots K)$ and $d^D \chi$ according to the order present  
in ${\cal A}$.
Additional factors (as $(i)^\frac{(D-1)D}{2}$ to $d^D \chi$ 
for extra insertions
of $\gamma_\chi$) will be resumed later on.
\par
Now we proceed to factor out the exponential $e^{\eta\xi}$
\begin{eqnarray}&&
Tr\Big( {\cal A}\gamma_\chi {\cal B} \gamma_\chi{\cal C}\Big)
=(-)^\frac{(D-1)D}{2}\int\, d{\cal C}\, d^D\xi\, d{\cal B}\, d^D \eta\,  d{\cal A}
\exp\Bigg({\cal C}*{\cal C}+ {\cal A}{\cal C} -  {\cal B}{\cal C}
\nonumber\\&&
+\big[-{\cal C}+{\cal B}+{\cal A}+\eta\big]
\big[\xi+{\cal C}+{\cal B}-{\cal A}\big] 
-{\cal A}{\cal B}
+{\cal B}*{\cal B}+{\cal A}*{\cal A}
\Bigg)
\label{gener.2.1}
\end{eqnarray}
Then we use the substitution principle and get
\begin{eqnarray}&&
Tr\Big( {\cal A}\gamma_\chi {\cal B} \gamma_\chi{\cal C}\Big)
=(i)^{(D-1)D}\int\,d{\cal C}\, d^D\xi\, d^D \eta\, d{\cal B}\,  d{\cal A}
\exp\bigg({(-)^{n_{\cal B}}\eta\xi}\bigg)
\nonumber\\&&
\exp\Bigg({\cal C}*{\cal C}+ {\cal A}{\cal C} -  {\cal B}{\cal C}
+{\cal B}*{\cal B}-  {\cal A} {\cal B}+{\cal A}*{\cal A}
\Bigg),
\label{gener.2.2}
\end{eqnarray}
where ${n_{\cal B}}$ is the number of switches $\eta_\mu \to -\eta_\mu$
necessary to bring $d^D \eta$ contiguous to $d^D \xi$ (i.e. the
number of elements of ${\cal B}$). Finally we 
change the sign ($p_j\to -p_j$) of all elements of ${\cal B}$ and get our final form
of the pairing process
\begin{eqnarray}&&
Tr\Big( {\cal A}\gamma_\chi {\cal B} \gamma_\chi{\cal C}\Big)
=(i)^{(D-1)D}\int\,d{\cal C}\, d^D\xi\, d^D \eta\, d{\cal B}\,  d{\cal A}
\exp\bigg({(-)^{n_{\cal B}}\eta\xi}\bigg)
\nonumber\\&&
\exp\Bigg({\cal C}*{\cal C}+ {\cal A}{\cal C} -  {\cal B}{\cal C}
+{\cal B}*{\cal B}- {\cal A} {\cal B}+{\cal A}*{\cal A}
\Bigg)
\nonumber\\&&
=(i)^{(D-1)D}\int\, d^D\xi\, d^D \eta\, 
\exp\bigg({(-)^{n_{\cal B}}\eta\xi}\bigg)
Tr\Big( {\cal A} \Big(-{\cal B}\Big) {\cal C}\Big).
\label{gener.2.3}
\end{eqnarray}
Eq. (\ref{gener.2.3}) shows that the  elements of ${\cal B}$ 
encapsulated between two $\gamma_\chi$ change sign in the process of pairing
and a $D$-dependent numerical factor is generated 
\begin{eqnarray}
\int\, d^D\xi\, d^D \eta\, 
\exp\bigg({(-)^{n_{\cal B}}\eta\xi}\bigg).
\label{gener.2.33}
\end{eqnarray}
This result is very surprising (and simple!). It allows to get rid of
all $\gamma_\chi$, but one if their number is odd. For sake of conciseness
$\gamma_\chi$ might enter only in $\cal C$.
\par\noindent Comments. i) For integer $D$ the eq. (\ref{more.2.2})
can be used and eq. (\ref{gener.2.3}) represents the algebra
in eq. (\ref{more.8.1.1}). ii) Pairing in eq. (\ref{gener.2.3}) for generic $D$  
is a generalization of the pairing for integer $D$ e.g. in eq. (\ref{more.8pp}).
%
{
\section{Cyclicity: General Proof }
\label{sec:cyclgen}
In the present Section we  prove that the trace integral
representation 
\begin{eqnarray}&& 
Tr(\not\! p_{1} \not\! p_2\dots\not \!p_k \gamma_\chi \not\! p_{k+1}\dots
\not\! p_{N} )
=
i^{\frac{D(D-1)}{2}} \int\,  d \bar c_{_N}\dots d\bar c_{_{k+1}}\, 
d^{^D}\xi\, d^{^{k}} \bar c\, 
\nonumber\\&& 
\exp \Big(\xi \big(\sum_{j=k+1}^{^N} p_{_j}\bar c_{_j}\big) 
+  \sum_{i=1}^{^k}\bar c_i  p_i\xi 
+\sum_{i<j=1}^N\bar c_i( p_{i}, p_{j}) \bar c_j\Big)
\label{cyclgen.1}
\end{eqnarray}
satisfies cyclicity
\begin{eqnarray}&& 
Tr(\not\! p_{1} \not\! p_2\dots\not \!p_k \gamma_\chi \not\! p_{k+1}\dots
\not\! p_{N} )
\nonumber\\&& 
=
Tr(\not\! p_{N}\not\! p_{1} \not\! p_2\dots\not \!p_k \gamma_\chi \not\! p_{k+1}\dots
\not\! p_{_{(N-1)}} ).
\label{cyclgen.2}
\end{eqnarray}
\underline{Proof}. Consider first $1<k<N$.
To accomplish the result we move $ d \bar c_{_N}$ to the far right thus
producing: i) a factor $(-)^{^{(N-1)}}$ and ii) a change $\xi \to -\xi$
in the Grassmannian variable which can be absorbed by a redefinition of
$\xi$.
\begin{eqnarray}&& 
Tr(\not\! p_{1} \not\! p_2\dots\not \!p_k \gamma_\chi \not\! p_{k+1}\dots
\not\! p_{N} )
=
i^{\frac{D(D-1)}{2}} \int\,  d \bar c_{_{(N-1)}}\dots d\bar c_{_{k+1}}\, 
d^{^D}\xi\, d^{^{k}} \bar c\, d \bar c_{_N}
\nonumber\\&& 
(-)^{^{(N-1)}}
\exp \Big(-\xi \big(\sum_{j=k+1}^{^N} p_{_j}\bar c_{_j}\big) 
-  \sum_{i=1}^{^k}\bar c_i  p_i\xi 
+\sum_{i<j=1}^N\bar c_i( p_{i}, p_{j}) \bar c_j\Big).
\label{cyclgen.3}
\end{eqnarray}
Now we change sign to the 
variables $\big(\bar c_{_j}, j=1,\dots,N-1\big)$
\begin{eqnarray}&& 
Tr(\not\! p_{1} \not\! p_2\dots\not \!p_k \gamma_\chi \not\! p_{k+1}\dots
\not\! p_{N} )
=
i^{\frac{D(D-1)}{2}} \int\,  d \bar c_{_{(N-1)}}\dots d\bar c_{_{k+1}}\, 
d^{^D}\xi\, d^{^{k}} \bar c\, d \bar c_{_N}
\nonumber\\&& 
\exp \Big(\xi \big(\sum_{j=k+1}^{^{N-1}} p_{_j}\bar c_{_j}\big) 
+  \sum_{i=1}^{^k}\bar c_i  p_i\xi + \bar c_{_N}p_{_N}\xi
+\sum_{i<j=1}^{^{N-1}}\bar c_i( p_{i}, p_{j}) \bar c_j
\nonumber\\&& 
-\sum_{i=1}^{^{N}}\bar c_{_i}( p_{_i}, p_{_{N}}) \bar c_{_{N}} \Big)
\label{cyclgen.4}
\end{eqnarray}
i.e.
\begin{eqnarray}&& 
Tr(\not\! p_{1} \not\! p_2\dots\not \!p_k \gamma_\chi \not\! p_{k+1}\dots
\not\! p_{N} )
=
i^{\frac{D(D-1)}{2}} \int\,  d \bar c_{_{(N-1)}}\dots d\bar c_{_{k+1}}\, 
d^{^D}\xi\, d^{^{k}} \bar c\, d \bar c_{_N}
\nonumber\\&& 
\exp \Big(\xi \big(\sum_{j=k+1}^{^{N-1}} p_{_j}\bar c_{_j}\big) 
+  \sum_{i=1}^{^k}\bar c_i  p_i\xi + \bar c_{_N}p_{_N}\xi
+\sum_{i<j=1}^{^{N-1}}\bar c_i( p_{i}, p_{j}) \bar c_j
\nonumber\\&& 
+\sum_{i=1}^{^{N}}\bar c_{_{N}}( p_{_{N}}, p_{_i}) \bar c_i \Big)
=Tr(\not\! p_{N}\not\! p_{1} \not\! p_2\dots\not \!p_k \gamma_\chi \not\! p_{k+1}\dots
\not\! p_{_{(N-1)}} ).
\label{cyclgen.5}
\end{eqnarray}
Eq. (\ref{cyclgen.5}) shows that cyclicity is valid in
generic dimensions if we move any gamma. 
\par
We have to prove that cyclicity is valid also if
$\gamma_\chi$ is moved around. 
}  
Start with
\begin{eqnarray}&& 
Tr(\not\! p_{1} \not\! p_2\dots \not\! p_{N-1} \not\! p_{N}\gamma_\chi )
=
i^{\frac{D(D-1)}{2}} \int\, d^{^D}\xi\,  d^{^{N}} \bar c\,  
\nonumber\\&& 
\exp \Big(  \sum_{i=1}^{N}\bar c_i  p_i \xi
+\sum_{i<j=1}^N\bar c_i( p_{i}, p_{j}) \bar c_j\Big)
\label{cycle.5.1}
\end{eqnarray}
and switch $d^{^D}\xi d^{^{N}}\bar c\to d^{^{N}}\bar c
d^{^D}\xi$. Due to the anticommutation rules
we get a factor $(-)^{^N}$ on each single differential $d\xi_\mu$.
\begin{eqnarray}&& 
Tr(\not\! p_{1} \not\! p_2\dots \not\! p_{N-1} \not\! p_{N}\gamma_\chi )
=
i^{\frac{D(D-1)}{2}} \int\, d^{^{N}} \bar c\, \prod_\mu d(-)^N\xi_\mu\,   
\nonumber\\&& 
\exp \Big(  \sum_{i=1}^{N}\bar c_i  p_i \xi
+\sum_{i<j=1}^N\bar c_i( p_{i}, p_{j}) \bar c_j\Big)
\label{cycle.5.1.1}
\end{eqnarray}
\par
We proceed by considering first $N$ even, i.e. $(-)^{^N}=1$.
Thus the change of variable $\bar c_j\to -\bar c_j , \forall j=1,\dots,N$
in eq. (\ref{cycle.5.1}) yields
\begin{eqnarray}&& 
Tr(\not\! p_{1} \not\! p_2\dots \not\! p_{N-1} \not\! p_{N}\gamma_\chi )
=
i^{\frac{D(D-1)}{2}} \int\,  d^{^{N}} \bar c\,  d^{^D}\xi\, 
\nonumber\\&& 
\exp \Big( - \sum_{i=1}^{N}\bar c_i  p_i \xi
+\sum_{i<j=1}^N\bar c_i( p_{i}, p_{j}) \bar c_j\Big)
\nonumber\\&& 
=
Tr(\gamma_\chi\not\! p_{1} \not\! p_2\dots \not\! p_{N-1} \not\! p_{N} ).
\label{cycle.5.11}
\end{eqnarray}
\par
For odd $N$ we have $(-)^{^N}=-1$ in eq. (\ref{cycle.5.1.1}), thus the change $\xi \to -\xi$
is needed in order to have the correct measure of the integral
\begin{eqnarray}&& 
Tr(\not\! p_{1} \not\! p_2\dots \not\! p_{N-1} \not\! p_{N}\gamma_\chi )
=
i^{\frac{D(D-1)}{2}} \int\,  d^{^{N}} \bar c\,  d^{^D}\xi\, 
\nonumber\\&& 
\exp \Big( - \sum_{i=1}^{N}\bar c_i  p_i \xi
+\sum_{i<j=1}^N\bar c_i( p_{i}, p_{j}) \bar c_j\Big)
\nonumber\\&& 
=
Tr(\gamma_\chi\not\! p_{1} \not\! p_2\dots \not\! p_{N-1} \not\! p_{N} ).
\label{cycle.5.12}
\end{eqnarray}
Eqs. (\ref{cycle.5.11}) and (\ref{cycle.5.12}) show that
cyclicity is obeyed also for $\gamma_\chi$.
\section{Charge Conjugation }
\label{sec:charge}
After we have established the validity of pairing in Section
\ref{sec:gener} and of cyclicity in Section \ref{sec:cyclgen}
we can discuss Charge Conjugation. We shall compare the algebra
 in the integral representation eqs. (\ref{intro.001a}) 
and (\ref{intro.001b}) 
and in the matrix representation
for the limit $D\to D_0$, where the integer $D_0$ is the  dimension
of the target space-time.

\subsection{Charge Conjugation in the Matrix representation}
%
We can assume that there is only one $\gamma_\chi$ present in the trace 
and  that the position of $\gamma_\chi$ is far right.
\par
In the matrix representation and for integer $D_0$ the following
properties are of common use
\begin{eqnarray}&& 
Tr\Big(\not\!p_1 \dots \not\!p_N\Big)
= 
Tr\Big(\big(\not\!p_1 \dots \not\!p_N\big)^T\Big)
= 
Tr\Big(\not\!p_N^T \dots \not\!p_1^T\Big)
\nonumber\\&&
Tr\Big(\not\!p_1 \dots \not\!p_k\gamma_\chi\dots\not\!p_N\Big)
= 
Tr\Big(\big(\not\!p_1 \dots \not\!p_k\gamma_\chi\dots\not\!p_N\big)^T\Big)
\nonumber\\&&
= 
Tr\Big(\big(\not\!p_N^T \dots\gamma_\chi^T \not\!p_k^T\dots\not\!p_1^T\big)\Big)
\label{charge.1}
\end{eqnarray}
A unitary Charge Conjugation operator $\cal C$ is  introduced
\begin{eqnarray}&& 
{\cal C}\gamma_\alpha {\cal C}^{-1} =- \gamma_\alpha ^T
\nonumber\\&&
{\cal C}\gamma_\chi {\cal C}^{-1} 
=(-)^{D_0 }\gamma_1^T\cdots \gamma_{D_0}^T
\nonumber\\&&
=(-)^{D_0 } (-)^{\frac{D_0(D_0-1)}{2}}\gamma_\chi ^T
=(-)^{\frac{D_0(D_0+1)}{2}}\gamma_\chi ^T
\nonumber\\&&
\label{charge.2}
\end{eqnarray}
and then 
\begin{eqnarray}&& 
Tr\Big(\not\!p_1 \dots \not\!p_N\Big)
=
Tr\Big(\big(\not\!p_1 \dots \not\!p_N\big)^T\Big)
=
Tr\Big(\not\!p_N^T \dots \not\!p_1^T\Big)=
\nonumber\\&&
(-)^NTr\Big({\cal C}\not\!p_N{\cal C}^{-1} \dots{\cal C} \not\!p_1{\cal C}^{-1}\Big)
=
(-)^NTr\Big(\not\!p_N \dots \not\!p_1\Big)
\nonumber\\&&
=
Tr\Big(\not\!p_N \dots \not\!p_1\Big)
\label{charge.3}
\end{eqnarray}
$N$ must be even because there are no $\gamma_\chi$ in the trace.
\par
If there is a $\gamma_\chi$ in the trace
\begin{eqnarray}
N = D_0 + 2n, \quad n\in {\cal N}
\label{charge.4}
\end{eqnarray}
From eqs. (\ref{charge.1}) and (\ref{charge.2}) we get

\begin{eqnarray}&& 
Tr\Big(\not\!p_1 \dots \not\!p_k\gamma_\chi\dots\not\!p_N\Big)
= 
Tr\Big(\big(\not\!p_1 \dots \not\!p_k\gamma_\chi\dots\not\!p_N\big)^T\Big)
\nonumber\\&&
= 
Tr\Big(\big(\not\!p_N^T \dots\gamma_\chi^T \not\!p_k^T\dots\not\!p_1^T\big)\Big)
\nonumber\\&&
=(-)^N (-)^{\frac{D_0(D_0+1)}{2}}
Tr\Big({\cal C}\not\!p_N{\cal C}^{-1} \dots{\cal C}\gamma_\chi{\cal C}^{-1}\dots
{\cal C} \not\!p_1{\cal C}^{-1}\Big)
\nonumber\\&&
=(-)^{D_0}(-)^{\frac{D_0(D_0+1)}{2}}
Tr\Big(\not\!p_N \dots\gamma_\chi\dots
 \not\!p_1\Big)
\nonumber\\&&
=(-)^{\frac{D_0(D_0-1)}{2}}
Tr\Big(\not\!p_N \dots\gamma_\chi\dots
 \not\!p_1\Big)
.
\label{charge.5}
\end{eqnarray}
In eqs. (\ref{charge.3}) and (\ref{charge.5}) the
$N$-dependent factors  have disappeared.
\subsection{Charge Conjugation in the Integral Representation }
The problem is now whether eqs.   (\ref{charge.3}) and (\ref{charge.5})
can be extended to the case of generic $D$  (while $D_0$ is kept fixed).
Let us see what  the integral representation gives for the relation 
(\ref{charge.3}). We consider
\begin{eqnarray}&& 
Tr\Big(\not\!p_N \dots \not\!p_1\Big)
=\int d\bar c_N\dots d\bar c_1 \exp\Big(\sum_{i<j}\bar c_i p_{N-i+1}p_{N-j+1}\bar c_j
\Big)\label{charge.6}
\end{eqnarray}
We redefine $\bar b_i =\bar c_{N-i+1}$. Then we  rearrange the measure of the integral
getting a factor 
$(-)^{\frac{N(N-1)}{2}}$ and the order in every pairing getting a factor $(-)^{\frac{N}{2}}$
\begin{eqnarray}&& 
Tr\Big(\not\!p_N \dots \not\!p_1\Big)
=(-)^{\frac{N(N-1)}{2}}(-)^{\frac{N}{2}}
Tr\Big(\not\!p_1 \dots \not\!p_N\Big)
\nonumber\\&&
=
Tr\Big(\not\!p_1 \dots \not\!p_N\Big),
\label{charge.7}
\end{eqnarray}
In fact  $N$ must be even for a trace with no $\gamma_\chi$. Eq. (\ref{charge.7})
is a consistent continuation of eq. (\ref{charge.3}) to 
generic value of $D$.
\par
Now we consider the trace with one $\gamma_\chi$ and get
\begin{eqnarray}&& 
Tr\Big(\gamma_\chi\not\!p_N \dots \not\!p_1\Big)
=
 \int d\bar c_N\dots d\bar c_1 \int d^{{D}}\xi \exp\Big(
\sum_i\xi \bar c_i p_{_{_{N-i+1}}}
\nonumber\\&&
+\sum_{i<j}\bar c_i p_{_{N-i+1}}p_{_{N-j+1}}\bar c_j
\Big)\label{charge.8}
\end{eqnarray}
The redefinition $\bar b_i =\bar c_{_{N-i+1}}$ gives 
\begin{eqnarray}&& 
Tr\Big(\gamma_\chi\not\!p_N \dots \not\!p_1\Big)
=
\int d^{D}\xi (-)^{\frac{N(N-1)}{2}}\int d\bar b_N\dots d\bar b_1 
\nonumber\\&&
\exp\Big(-
(-)^N \sum_i \bar b_i p_{i}\xi
-\sum_{i<j}\bar b_i p_{i}p_{j}\bar b_j
\Big)\label{charge.9}
\end{eqnarray}
where the factor $(-)^{\frac{N(N-1)}{2}}$ comes from rearrangement
of $ d\bar b_1\dots d\bar b_N$ and the change of variables
$\xi_\mu \to (-)^N \xi_\mu$ is performed in order to 
get rid of the parity factor in the switch of $d^N\bar b d^D\xi
\to   d^D\xi  d^N\bar b  $.
Let $D_0 $ be the number of $ \bar b p_i\xi$ present in the
integral over $\xi$. $D_0$ is an integer. $\frac{N-D_0}{2}$
is the number of pairs in the integrand  of  $d^N \bar b$. There
is no analytic continuation on $D_0$ and it is fixed by $N$ and the
number of $\bar b$ entering in the $\bar b_i p_i p_j \bar b_j$
terms. Thus after the expansion of the exponential 
eq. (\ref{charge.9}) becomes
\begin{eqnarray}&& 
Tr\Big(\gamma_\chi\not\!p_N \dots \not\!p_1\Big)
=\sum_{\cal P}
\int d^{D}\xi (-)^{\frac{N(N-1)}{2}}\int d\bar b_N\dots d\bar b_1 
\nonumber\\&&
(-)^{(N+1)D_0}\prod_{i_p\in{\cal P} } \bar b_{i_p } p_{{i_p}}\xi
(-)^{\frac{N-D_0}{2}}
\prod_{k<l \not\in {\cal P} }\bar b_k p_{k}p_{l}\bar b_l
\label{charge.10}
\end{eqnarray}
where ${\cal P}$ is an oriented set of $D_0$ positive integers
$<N.$ 
\par
We use the relation
\begin{eqnarray}&&
N=D_0 ~~{\rm Mod}(2)
\label{charge.10p}
\end{eqnarray}
to get
\begin{eqnarray}&&
(-)^{(N+1)D_0   }
=
(-)^{  (N-D_0)D_0 + D_0^2 + D_0  }
\nonumber\\&&
 =(-)^{   D_0(D_0 + 1) } =1
\label{charge.10pp}
\end{eqnarray}
and 
\begin{eqnarray}&&
(-)^\frac{N(N-1)}{2}
\nonumber\\&&
=
(-)^{\frac{(N-D_0)^2}{2}+ \frac{(N-D_0)D_0}{2} +\frac{(N-D_0)(D_0-1)}{2}+ \frac{(D_0-1)D_0}{2}   }
\nonumber\\&&
=
(-)^{  -\frac{(N-D_0)}{2}+ \frac{(D_0-1)D_0}{2}   }
\label{charge.10q}
\end{eqnarray}
The resulting factor gives the final expression of eq. (\ref{charge.9})
\begin{eqnarray}&& 
Tr\Big(\gamma_\chi\not\!p_N \dots \not\!p_1\Big)
=
(-1)^{\frac{D_0(D_0-1)}{2}}
Tr\Big(\not\!p_1 \dots \not\!p_N\gamma_\chi\Big)
\label{charge.11}
\end{eqnarray}
which agrees with eq. (\ref{charge.5}).
\par
Eq. (\ref{charge.11}) is the important result of the Section. One
can perform the transposed of the argument of the trace and subsequently
get rid of the transposition on the single gamma matrices by using
the matrix $\cal C$. The result is the reversed order of the momenta and the
switch ($\gamma_\alpha \gamma_\chi \to \gamma_\chi\gamma_\alpha $).
$D_0$ is an initial data of the problem. The Laurent expansion
is around $D_0$. 
\section{Subtraction Strategy}
\label{sec:str}
We can separate the integration over the variables generating the
$\gamma_\chi$ from those producing the trace without $\gamma_\chi$
denoted by $\bar c$. This splitting is quite natural if one
uses our integral representation of the trace. The new 
Regularization procedure 
consists in operating the necessary pole subtractions
on the integrand, where the dimension is not fixed to integer
values.
\par
The new renormalization procedure is described by the following 
assumption
\par\noindent
{\bf Assumption 2:} 
\par\noindent {\sl
``The Dimensional subtractions have to be 
performed \underline{before} the final integration on 
the Grassmannian variable that produces $\gamma_\chi$ in the
trace (i.e. $d^{^D}\xi$ in eq. (\ref{intro.001b})).''}
\par
For one $\gamma_\chi$ the integrand is 
\begin{eqnarray}&&
R(p_1\dots p_N |\xi)\equiv
 \int\, d^{^N} \bar c\, 
\exp \Big(  \sum_{i=1}^{N}\bar c_i  (p_i)_\mu\xi_\mu 
+\sum_{i<j=1}^N\bar c_i( p_{i}, p_{j}) \bar c_j\Big),
\label{intro.001b.0}
\end{eqnarray}
i.e. we drop the integration over $d^D\xi$ and consequently it
becomes function of the Grassmannian real 
variable $\xi$. The bar $|$ separates variable $\xi$ form $p_i$'. The
ordered sequence is dictated by the order in the exponential and in the
measure on $\bar c$ of eq. (\ref{intro.001b.0} )
\par
According to the above prescription,
after 
{ all poles  in $D$ have been subtracted,} 
the limit on $D$ to the required integer is performed by the final 
integration on $ 
{i^{\frac{D(D-1)}{2}}} \int d^D \xi$ in eq. (\ref{intro.001b.0}). 
\par\noindent
{\bf Comment: it should be stressed that Assumption 2 allows
computer algebraic manipulations, since the algebra does not
depend on the dimensions.}

\section{Strategy in  Manipulation of $\gamma_\chi$}
\label{sec:prop}
In this paper we propose a simple strategy in order to deal with
the pole subtraction. We will avoid generality in the procedure
in favor of a simple set of rules, easy to implement.
\begin{enumerate}
\item It is assumed that the substitution principle
($\xi \to \xi + {\rm constant}$) is allowed in the integration.
Moreover the change of integration variables is performed in 
the standard way by using the Jacobian.
\item By the method of pairing all $\gamma_\chi$ are removed
by proceeding from the far left.
\item The last $\gamma_\chi$ is eventually brought to the far 
right of the trace by using cyclicity.
\item The poles subtraction is performed on the final expression obtained
from the above enumerated rules.
\item Integration $\int d^D\xi$ over the  Grassmannian variable
$\xi$ is performed after pole subtraction.
\end{enumerate}
%
\section{The Case of a Single \huge{$\gamma_\chi$}}
\label{sec:sin}
By using pairing as in eq. (\ref{gener.2.3}) all $\gamma_\chi$
can be removed and the
trace contains at most one $\gamma_\chi$. Moreover
cyclicity allows to bring it to the far right.
Thus we have the trace in canonical form as in eq.  
(\ref{intro.001b}):
\begin{eqnarray}&&
Tr(\not\! p_{1} \not\! p_2\dots \not\! p_{N-1} \not\! p_{N}\gamma_\chi )
=
i^{\frac{D(D-1)}{2}}  \int d^{^D}\xi\, d^{^N} \bar c\, 
\exp \Big(  \sum_{i=1}^{N}\bar c_i  (p_i)\xi
\nonumber\\&& 
+\sum_{i<j=1}^N\bar c_i( p_{i}, p_{j}) \bar c_j\Big).
\label{intro.001bb}
\end{eqnarray}
On the integrand of $d^D\xi$ we perform all the
necessary subtractions by using the standard procedure
based on the use of local counter-terms in the action
\begin{eqnarray}
{\cal S} \to \sum_{n=0} {\cal S}^{n}
\label{sin.2}
\end{eqnarray}
expanded in the number of loops. For instance, if a counterterm
is necessary,
at one loop for the chiral current insertion could be 
\begin{eqnarray}
{\cal S}^{1}\equiv \int d^D x
\frac{J_\alpha}{D-4}\Big( A\bar\psi\gamma_\alpha \gamma_\chi\psi
+B \bar\psi\gamma_\alpha \gamma_\tau\gamma_\chi\gamma_\tau\psi
+C\bar\psi\gamma_\alpha \gamma_\iota\gamma_\tau\gamma_\chi\gamma_\tau
\gamma_\iota\psi+\dots\Big).
\label{sin.3}
\end{eqnarray}
Notice that all terms in eq. (\ref{sin.3}) are equivalent
since 
\begin{eqnarray}
\{\gamma_\chi,\gamma_\tau\}\sim {\cal O}(D-4).
\label{sin.4}
\end{eqnarray}
The choice of $A,B,C$ is dictated by some extra arguments
(symmetry, topological arguments, etc.).
\par
In the aftermath we have to evaluate the integral $d^D\xi$
in eq. (\ref{intro.001bb})
at some integer value. Say $D=4$. This implies that we
have to collect all the monomials with $\xi$ at the fourth power
containing all $\bar c_j ,  j=1,\dots, N$. $\bar c_j$ might
originate from the expansion of
\begin{eqnarray}
 \exp\sum_{i<j} \bar c_j (p_j, p_{j})\bar c_{j} \sim 
\prod_{i<j}\left[1 + \bar c_j (p_j, p_{j})\bar c_{j}\right]
\label{sin.5}
\end{eqnarray}
or from the expansion
\begin{eqnarray}
 \exp \sum_j\bar c_j (p_j, \xi) \sim \prod_j\left[1 + \bar c_j (p_j, \xi)\right].
\label{sin.6}
\end{eqnarray}
The momenta of eq. (\ref{sin.6}) will saturate the
completely antisymmetric tensor. The rest of momenta
taken from the expansion (\ref{sin.5}) will yield the
usual trace without $\gamma_\chi$ as in eq. (\ref{intro.001a}).
\subsection{General Properties of the Trace}
\label{sec:gen}
By general properties  we denote those valid also
when the integration over $d^D\xi$ is suspended because
$D$ is not given as an integer.
\par
The rules (\ref{intro.002}) of Grassmann integration requires that
the single $\bar c_j$  appears only once
in the integrands of eq. (\ref{intro.001b.0}). Then it
is a linear function of $p_j$, for each $j=1,\dots,N$. 
Thus we can expand it in powers of $\xi$ by using
\begin{eqnarray}&&
\exp \Big(  \sum_{i=1}^{N}\bar c_i  (p_i)\xi\Big)
= \prod_{i=1}^N  e^{\bar c_i  (p_i\xi)}
= \prod_{i=1}^N (1+ \bar c_i  (p_i\xi))
\nonumber\\&& 
= 1 + \sum_{i=1}^N\bar c_i  (p_i\xi) +
\sum_{i,j=1}^N\bar c_i  (p_i\xi)\bar c_j  (p_j\xi) +
\sum_{i,j,k=1}^N\bar c_i  (p_i\xi)\bar c_j  (p_j\xi)\bar c_k  (p_k\xi)
\nonumber\\&& 
+\dots.
\label{intro.001b.1}
\end{eqnarray}
\par
The integration on $d^N \bar c$ is partly on the chosen 
terms of the expansion (\ref{intro.001b.1}) and the rest
on the Taylor expansion of the other exponential factor
in eq. (\ref{intro.001b.0}). These last terms
are bilinear in $\bar c$, thus the power of $\xi$
is even or odd depending on the value of $mod(N,2)=0$ or
$mod(N,2)=1$.

\subsection{Fundamental Formula}
After the integration  on $\bar c$ generated by the expansion in
eq. (\ref{intro.001b.1}) a typical expression in terms
of powers of $\xi$ is given by eq. (\ref{intro.001bb})
\begin{eqnarray}&&
 \int d^{^N} \bar c\, 
\exp \Big(  \sum_{i=1}^{N}\bar c_i  (p_i)\xi
+\sum_{i<j=1}^N\bar c_i( p_{i}, p_{j}) \bar c_j\Big)
\nonumber\\&& 
=
\sum_{\cal P} \int d\bar c_{j_K}\cdots d\bar c_{j_1}\delta_{\cal P}
(\bar c_{j_1}p_{j_1}\xi)\cdots (\bar c_{j_K}p_{j_K}\xi)
\nonumber\\&& 
Tr(\not\! p_{i_1} \dots\not\! p_{i_{N-K}} )
\nonumber\\&&
\label{intro.001b.2}
\end{eqnarray}
where the sum is over all partitions ${\cal P}$ of the
$N$ integers in two mutually disjoint ordered sets $(i_1,\dots,i_{N-K})$ 
and $(j_1,\dots, j_{K})$. The parity $\delta_{\cal P}$
counts the permutations needed to perform the integrations
over $d\bar c_{j_K},\dots, d\bar c_{j_1} d\bar c_{i_{(N-K)}},\dots, d\bar c_{i_1} 
$. The integration over
$\bar c$ yields
\begin{eqnarray}&&
 \int d^{^N} \bar c\, 
\exp \Big(  \sum_{i=1}^{N}\bar c_i  (p_i)\xi
+\sum_{i<j=1}^N\bar c_i( p_{i}, p_{j}) \bar c_j\Big)
\nonumber\\&& 
=
\sum_{\cal P}\delta_{\cal P} ( p_{j_K}\xi)\dots
(p_{j_1}\xi)
Tr(\not\! p_{i_1} \dots\not\! p_{i_{N-K}} )
\nonumber\\&&
\label{intro.001b.3}
\end{eqnarray}
The quantity in eq. (\ref{intro.001b.3}) is the perfect tool
to be continued in $D$. It has simple symmetry properties:
by interchanging two adjacent $p_j$'s we get a minus sign.

\begin{eqnarray}&&
\sum_{\cal P}\delta_{\cal P} ( p_{j_K}\xi)\dots
(p_{j_{(n+1)}}\xi)
(p_{j_n}\xi)\dots
(p_{j_1}\xi)
Tr(\not\! p_{i_1} \dots\not\! p_{i_{N-K}} ) 
\nonumber\\&&
= -
\sum_{\cal P}\delta_{\cal P} ( p_{j_K}\xi)\dots
(p_{j_n}\xi)(p_{j_{n+1}}\xi)\dots
(p_{j_1}\xi)
Tr(\not\! p_{i_1} \dots\not\! p_{i_{N-K}} ) 
\label{intro.001b.4}
\end{eqnarray}
If any two $(p_j\xi)$'s are equal, the quantity in eq. (\ref{intro.001b.3})is  zero.
%
\subsection{Clifford algebra}
%
\label{sec:cliff}
We now comment on how the Clifford Algebra is implemented
in eq. (\ref{intro.001b.2}). We follow closely the proof
given in Section 3 of Ref. \cite{Ferrari:2014jqa}. We write
in full evidence the variables  $p_n$ and $p_{n+1}$.
\begin{eqnarray}&&
R(p_1\dots p_N|\xi)=
 \int d^{^N} \bar c\, 
\exp \Big( 
\bar c_n p_{n}\xi + \bar c_{n+1} p_{n+1}\xi
+ \sum_{i\not =n,\not =n+1 }^{N}\bar c_i  (p_i)\xi\Big)
\nonumber\\&& 
\exp \Big(\bar c_n( p_n, p_{n+1}) \bar c_{n+1}
+
\sum_{j>n+1}^N\bar c_n( p_n, p_{j}) \bar c_j
+ 
\sum_{j>n+1}^N\bar c_{n+1}( p_{n+1}, p_{j}) \bar c_j
\nonumber\\&& 
+
\sum_{j<n}^N \bar c_j( p_{j}, p_n)\bar c_n
+ 
\sum_{j<n}^N\bar c_j( p_j, p_{n+1}) \bar c_{n+1}
+
\sum_{i,j\not=n,n+1,i<j}^N\bar c_i( p_{i}, p_{j}) \bar c_j
\Big)
\label{intro.001b.6}
\end{eqnarray}
Now we  replace $p_n\leftrightarrow p_{n+1}$
\begin{eqnarray}&&
R(p_1\dots p_{n+1}p_n\dots p_N|\xi )
=
 \int d^{^N} \bar c\, 
\exp \Big( 
\bar c_n p_{n+1}\xi + \bar c_{n+1} p_{n}\xi
+ \sum_{i\not =n,\not =n+1 }^{N}\bar c_i  (p_i)\xi\Big)
\nonumber\\&& 
\exp \Big(\bar c_n( p_{n+1}, p_{n}) \bar c_{n+1}
+
\sum_{j>n+1}^N\bar c_n( p_{n+1}, p_{j}) \bar c_j
+ 
\sum_{j>n+1}^N\bar c_{n+1}( p_{n}, p_{j}) \bar c_j
\nonumber\\&& 
+
\sum_{j<n}^N \bar c_j( p_{j}, p_{n+1})\bar c_n
+ 
\sum_{j<n}^N\bar c_j( p_j, p_{n}) \bar c_{n+1}
+
\sum_{i,j\not=n,n+1,i<j}^N\bar c_i( p_{i}, p_{j}) \bar c_j
\Big)
\label{intro.001b.6p}
\end{eqnarray}
and rename $\bar c_n \leftrightarrow \bar c_{n+1}$
\begin{eqnarray}&&
=
- \int d^{^N} \bar c\, 
\exp \Big(  \sum_{i}^{N}\bar c_i  (p_i)\xi\Big)
\nonumber\\&& 
\exp \Big(\bar c_{n+1}( p_{n+1}, p_{n}) \bar c_{n}
+
\sum_{j>n+1}^N\bar c_{n+1}( p_{n+1}, p_{j}) \bar c_j
+ 
\sum_{j>n+1}^N\bar c_n( p_{n}, p_{j}) \bar c_j
\nonumber\\&& 
+
\sum_{j<n}^N\bar c_j( p_{j}, p_{n+1}) \bar c_{n+1}
+
\sum_{j<n}^N\bar c_j( p_j, p_{n}) \bar c_{n}
+
\sum_{i,j\not=n,n+1,i<j}^N\bar c_i( p_{i}, p_{j}) \bar c_j
\Big)
\nonumber\\&& 
\label{intro.001b.7p}
\end{eqnarray}
and finally
\begin{eqnarray}&&
R(p_1\dots p_{n+1}p_n\dots p_N|\xi )
=
- \int d^{^N} \bar c\, 
\exp \Big(  \sum_{i}^{N}\bar c_i  (p_i)\xi\Big)
\nonumber\\&& 
\exp (-2 \bar c_{n} ( p_{n}, p_{n+1})\bar c_{n+1})
\nonumber\\&& 
\exp \Big(\bar c_{n} ( p_{n}, p_{n+1})\bar c_{n+1}
+
\sum_{j>n+1}^N\bar c_{n+1}( p_{n+1}, p_{j}) \bar c_j
+ 
\sum_{j>n+1}^N\bar c_n( p_{n}, p_{j}) \bar c_j
\nonumber\\&& 
+
\sum_{j<n}^N\bar c_j( p_{j}, p_{n+1}) \bar c_{n+1}
+
\sum_{j<n}^N\bar c_j( p_j, p_{n}) \bar c_{n}
+
\sum_{i,j\not=n,n+1,i<j}^N\bar c_i( p_{i}, p_{j}) \bar c_j
\Big)
\nonumber\\&& 
\label{intro.001b.7pp}
\end{eqnarray}
We expand the exponential and get
\begin{eqnarray}&&
R(p_1\dots p_{n}p_{n+1}\dots p_N|\xi )+
R(p_1\dots p_{n+1}p_n\dots p_N|\xi )
\nonumber\\&& 
= 2 ( p_{n}, p_{n+1}) 
R(p_1\dots p_{n-1}p_{n+2}\dots p_N|\xi ).
\label{intro.001b.7ppp}
\end{eqnarray}
The above equation
shows that eq. (\ref{intro.001b.2}) gives a representation
of the Clifford algebra. 


\subsection{Comparison with Previous Approaches}
%
\label{sec:comp}
In some points our approach resembles similar attempts:
where the algebra of $\gamma_\chi$ is never used up to
the very end of the calculation and it is finally resumed by some 
{\sl ``obvious''}
statements \cite{Ferrari:2014jqa},\cite{El-Menoufi:2015dra}.
For instance if $D=4$ is the final goal
then in the vicinity of $D=4$ one uses
\begin{eqnarray}&&
Tr(\gamma_\chi \gamma_\mu \gamma_\nu) = 0
\label{foul.1}
\end{eqnarray}
%
{
since the LHS of (\ref{foul.1}) is non-zero only for $D= 2$, i.e.
``too far away'' from the target $D=4$ value.}
\par
Unfortunately this procedure might cause inconsistencies.
For instance the use of the Clifford algebra gives ($\widehat{\cdots}$ means omitted)
\begin{eqnarray}&&
Tr(\gamma_\chi \gamma_{\mu_1}\cdots\gamma_{\mu_{j}}\gamma_{\mu_{j+1}} \cdots \gamma_{\mu_4} ) 
=-Tr(\gamma_\chi\gamma_{\mu_1} \cdots\gamma_{\mu_{j+1}}\gamma_{\mu_{j}} \cdots\gamma_{\mu_4})
\nonumber\\&&
+2 \delta_{\mu_{j}\mu_{j+1}}Tr(\gamma_\chi \gamma_{\mu_1} \cdots\widehat{ \gamma_{\mu_j}}
\widehat{ \gamma_{\mu_{j+1}}} \cdots \gamma_{\mu_4}).
\label{foul.2}
\end{eqnarray}
By using  eq. (\ref{foul.1}) one might conclude that the last term in eq. (\ref{foul.2})
is zero and then
\begin{eqnarray}&&
Tr(\gamma_\chi \gamma_{\mu_1}\cdots\gamma_{\mu_{j}}\gamma_{\mu_{j+1}} \cdots \gamma_{\mu_4} ) 
=-Tr(\gamma_\chi\gamma_{\mu_1} \cdots\gamma_{\mu_{j+1}}\gamma_{\mu_{j}} \cdots\gamma_{\mu_4})
\label{foul.3}
\end{eqnarray}
By repeated use of (\ref{foul.3}) and 
 by using cyclicity as in Section \ref{sec:cyclgen} one gets 
\begin{eqnarray}&&
Tr(\gamma_\chi  \gamma_{\mu_1}\cdots \gamma_{\mu_4})
=-Tr(  \gamma_{\mu_1}\gamma_\chi\cdots \gamma_{\mu_4})
\label{foul.4}
\end{eqnarray}
i.e. $\gamma_\chi$ is anticommuting.  In the trace with four or less
gamma's the algebra is the na\"{\i}ve one. Then well known relations
can be derived
\begin{eqnarray}&&
(D-4)(D-2)Tr(\gamma_\chi  \gamma_{\mu_1}\cdots \gamma_{\mu_4})=0,
\label{foul.5}
\end{eqnarray}
which conflicts with smooth continuation near $D\sim 4$.

The same objection
 does not apply to eq. (\ref{intro.001b.3}) because
$\gamma_\chi$ is not present until the final $\xi$ integration 
is performed.
\section{Ward Identity}
\label{sec:ward}
%
%
It is very interesting to see what is the destiny of
Ward identities in the present formalism. In fact the use
of DR guarantees that standard algebraic manipulations become
possible, since undefined mathematical objects do not show up any more.
On the other side the presence of $\gamma_\chi$ modifies 
the algebra of $\gamma$'s.
\par
We first consider the tree level at $D=4$
\begin{eqnarray}&&
(p- k)_\mu \gamma_\mu \gamma_5
\nonumber\\&&
=({\not\!p-m})\gamma_5 +  \gamma_5({\not\! k-m})
+ 2m  \gamma_5.
\label{ward.1}
\end{eqnarray}
By this Ward identity it is possible to connect  the vertex functions to
the two-point functions. Divergences are  related in the
corresponding fashion.
$m$ is the explicit  chiral breaking parameter.
Similarly,
in generic $D\sim 4$ dimensions, we have to compare the traces 
{
\begin{eqnarray}&&
(p+k)_\mu Tr\Big({\cal A} \gamma_\mu \gamma_\chi \Big) 
=
Tr\Big({\cal A}  \not\! p\gamma_\chi\Big) -
 Tr\Big({\cal A} \gamma_\chi \not\! k\Big)
+k_\mu
Tr\Big({\cal A} \{\gamma_\mu,\gamma_\chi\} \Big). 
\nonumber\\&&
\label{ward.2b}
\end{eqnarray}
}
where ${\cal A}= \not\!p_2\not\!p_3\dots \not\! p_{_N}$.
We use the result in Ref. \cite{Ferrari:2014jqa}    Section  4
\begin{itemize}
\item for even $N$ (i.e. $D$ even if integer)
\begin{eqnarray}&&
Tr(\not\!p_2\not\!p_3\dots \not\! p_{_N}{\{\gamma_\chi,\not\! k\}})
=
Tr(\{\not\! k,\not\!p_2\not\!p_3\dots \not\! p_{_N}\}{\gamma_\chi})
\nonumber\\&&
= 2 \sum_{j=2}^N(-)^j(k,p_j)
Tr(\widehat{\not\! k}\dots\widehat{\not\!p_j}\dots \not\! p_{_N}
{\gamma_\chi})
\label{int.repr.1.9.2}
\end{eqnarray}
\item for odd $N$ (i.e. $D$ odd if integer)
\begin{eqnarray}&&
Tr(\not\!p_2\not\!p_3\dots \not\! p_{_N}{[\gamma_\chi,\not\! k]})
=
Tr([\not\! k,\not\!p_2\not\!p_3\dots \not\! p_{_N}]{\gamma_\chi})
\nonumber\\&&
= 2 \sum_{j=2}^N(-)^j(kp_j)
Tr{(}\widehat{\not\! k}\dots\widehat{\not\!p_j}\dots \not\! p_{_N}{\gamma_\chi}).
\label{int.repr.1.9.3}
\end{eqnarray}
\end{itemize}
The Ward identity for a closed line of fermions can be easily
derived.
\par
For even $N$ eq. (\ref{int.repr.1.9.2}) yields
{
\begin{eqnarray}&&
(p+k)_\mu Tr\big({\cal A}\gamma_\mu \gamma_\chi\big)
=
 Tr\big({\cal A}\not\! p \gamma_\chi\big)
- Tr\big({\cal A}\gamma_\chi\not\! k \big) 
+Tr\big(\{\not\! k,{\cal A}\}\gamma_\chi \big)
\nonumber\\&&
= Tr\big({\cal A}\not\! p \gamma_\chi\big)
- Tr\big({\cal A}\gamma_\chi\not\! k \big) 
\nonumber\\&&
+2 \sum_{j=2}^N(-)^j(k,p_j)
Tr(\dots\widehat{\not\!p_j}\dots \not\! p_{_N}
{\gamma_\chi})
\label{even}
\end{eqnarray}
}
\par
For odd $N$ eq. (\ref{int.repr.1.9.2}) yields
\begin{eqnarray}&&
(p+k)_\mu Tr\big({\cal A}\gamma_\mu \gamma_\chi\big)
=
Tr\big({\cal A}\not\!p \gamma_\chi\big)
+
 Tr\big({\cal A}\gamma_\chi \not\! k \big)
- Tr\big([\not\! k,{\cal A}]\gamma_\chi  \big)
\nonumber\\&&
=
 Tr\big({\cal A}\not\!p \gamma_\chi\big)
+
 Tr\big({\cal A}\gamma_\chi \not\! k \big)
-2 \sum_{j=2}^N(-)^j(k,p_j)
Tr(\dots\widehat{\not\!p_j}\dots \not\! p_{_N}
{\gamma_\chi}).
\nonumber\\&&
\label{odd}
\end{eqnarray}
The last term in both equations (\ref{even}) and (\ref{odd})
are vanishing for integer $D$: for even or odd values respectively.
\par
The Ward identities (\ref{even}) and (\ref{odd}) are very useful in perturbative
calculations, since in the RHS the first two terms together and the third carry a factor
$(D-D_0)$ ($D_0$ is the target value)  which annihilates all finite terms
of the loop integration: 
only the divergent terms are important. This fact simplifies the loop expansion calculations.
\par
The Ward identity (for instance eq. (\ref{ward.2b})) is very useful.
In fact in the case of classical chiral invariance ($D\to 4$) the LHS
is zero in the limit. Consequently also the RHS is zero. The last term
is zero because $\{\gamma_\mu,\gamma_\chi\}=0$ thus also the rest of the RHS
must carry a (D-4) factor. In the evaluation of the loop expansion
only the poles survive being every term multiplied by the factor
$(D-4)$. This strategy will be illustrated in Section \ref{App:WARD}
in the one-loop  ABJ anomaly. It will be used  in the 
evaluation of the two-loop corrections in a future work.
\section{Conclusions: the Rules}
\label{sec:rules}
In Part I of the present paper we have derived the algebra for the integral 
representation of the trace. We obtained the standard
algebra of the matrix representation of the gamma's
for  generic \underline{integer} $D$ dimensions. 
 $\gamma_\chi$ obeys  a consistent
algebra
\begin{eqnarray}&&
\gamma_\chi\gamma_\mu + (-)^D \gamma_\mu\gamma_\chi=0
\nonumber\\&&
\gamma_\chi^2 =1.
\label{rules.1}
\end{eqnarray}
The above algebra allows the use of pairing technique:
all pairs of $\gamma_\chi$ can be removed inside the trace.
\par
Lorentz covariance and cyclicity are properties of the trace
in the integral representation. The Clifford algebra of
the gamma's is also implemented.
\par
However this algebra cannot be continued to complex
$D$ since eq. (\ref{rules.1}) requires $[-(-)^D]^2 = 1$.
\par
In Part II we have presented a new way to remove 
this obstacle to the continuation in $D$ of the trace with 
$\gamma_\chi$. 
{
First we proved that in any generic $D$ the pairing is allowed
(in a weak form, since a $D$-dependent numerical factor remains
see eq. (\ref{gener.2.3})).
Thus we could  remove all possible $\gamma_\chi$. Then we used cyclicity
in order to write the trace in the canonical form i.e.
with the surviving $\gamma_\chi$ shifted to the far right. It should
be stressed that in general  $\gamma_\chi^2 \not  = 1$.
\par
Once the trace is written in this form,  
we have
considered the integrand of the Grassmannian 
variables $\xi$ generating $\gamma_\chi$ upon integration. 
The integrand is function of the momenta $p_j, j=1,\dots,N$ and 
of the Grassmannian real variables $\xi_\mu$.
\par
The integrand can be easily expanded  in a sum of  monomials 
\begin{eqnarray}
\left[
c_{j_1}(p_{j_1},\xi) \dots c_{j_K} (p_{j_K},\xi)
\right]
\label{rules.2}
\end{eqnarray}
for any choice  $\{j_1,\dots,j_{K}\}$ of $1,\dots,N$. In this
way we get rid of the completely antisymmetric tensor and only powers
of the momenta $(p_j,p_{j'})$ and $(p_j,\xi)$ survive. 
Thus Feynman amplitudes can be
evaluated in generic $D$ dimensions. 
\par
The new strategy consists in performing
the pole subtraction on the integrand.
\par
After the amplitudes are properly  defined for the required
\underline{integer} $D$ dimensions (by poles subtraction), the relevant integration over
the variables $\xi$ restores the completely antisymmetric
tensor. In eq. (\ref{rules.2}) only $K=D$ survive since the integration
is over $d^{^D}\xi$ with integer $D$.
}

\section*{Acknowledgements}
%
The warm and stimulating hospitality of CTP-MIT
is gratefully acknowledged. Part of this work has been
done at the Department of Physics of the University of Pisa
and of the INFN, Sezione di Pisa, to which he is very grateful.

\appendix

\section{Example of Subtraction: ABJ Anomaly}
\label{sec:ABJ}
Now we can try to use eq. (\ref{intro.001b.2}). 
Our procedure of pole subtraction removes any ambiguity
in Dimensional Renormalization. In particular, even in the
case where the \underline{sum} of all graphs at given loop order
is finite thanks to the presence of a $(D-4)$ factor
removing the pole, the divergent single graphs can be
manipulated in a safe way under the protection of
the Regularization.
\par
We consider the ABJ anomaly. The relevant term to be evaluated
is the divergent part of the Feynman amplitude in Ref. \cite{Ferrari:2014jqa}.
In particular we start from eq. (65) containing the trace factor
\begin{eqnarray}&&
\int_0^1 dx\int_0^x dy \frac{d^Dq}{(2\pi)^D}
[q^2 + k^2 y +p^2x-p^2y -(ky-px+py)^2]^{-3}
\nonumber\\&&
Tr\Big (\gamma_\mu \gamma_\alpha\gamma_\rho\gamma_\beta \gamma_\sigma\gamma_\iota
\gamma_\chi \Big)
b_\mu(q+r-k)_\alpha \epsilon_{1\rho}(q+r)_\beta\epsilon_{2\sigma}(q+r+p)_\iota.
\label{abj.1}
\end{eqnarray}
where $k,p$ are incoming momenta, $\epsilon_{1,2}$ the abelian field 
polarizations, $b_\mu$ an external source and $r=yk-xp + yp$ for
the Feynman parameters $x,y$.
Three terms are divergent
\begin{eqnarray}&&
Tr\Big (\gamma_\mu \gamma_\alpha\gamma_\rho\gamma_\beta \gamma_\sigma\gamma_\iota
\gamma_\chi \Big)
b_\mu(q)_\alpha \epsilon_{1\rho}(q)_\beta\epsilon_{2\sigma}(r+p)_\iota,
\nonumber\\&&
Tr\Big (\gamma_\mu \gamma_\alpha\gamma_\rho\gamma_\beta \gamma_\sigma\gamma_\iota
\gamma_\chi \Big)
b_\mu(r-k)_\alpha \epsilon_{1\rho}(q)_\beta\epsilon_{2\sigma}(q)_\iota,
\nonumber\\&&
Tr\Big (\gamma_\mu \gamma_\alpha\gamma_\rho\gamma_\beta \gamma_\sigma\gamma_\iota
\gamma_\chi \Big)
b_\mu(q)_\alpha \epsilon_{1\rho}(r)_\beta\epsilon_{2\sigma}(q)_\iota
.
\label{abj.2}
\end{eqnarray}
Now we have to expand all three expressions of eq. (\ref{abj.2})
according to eq. (\ref{intro.001b.2}). Of the many terms only those with
two $q$ in the $\xi$ factor are zero by symmetry. We can avoid this
lengthy procedure by using the symmetric integration in $q$ 
before we use the expansion of eq. (\ref{intro.001b.2})
\begin{eqnarray}&&
\gamma_\alpha \gamma_\rho \gamma_\alpha = (2-D)\gamma_\rho
\nonumber \\&&
\gamma_\alpha \gamma_\rho \gamma_\sigma \gamma_\alpha =
(D-4)\gamma_\rho \gamma_\sigma + 4 \delta_{\rho\sigma}
\nonumber \\&&
\gamma_\alpha \gamma_\rho \gamma_\beta\gamma_\sigma \gamma_\alpha =
(6-D)\gamma_\rho \gamma_\beta\gamma_\sigma 
\nonumber \\&&
-4(\delta_{\rho\beta}\gamma_\sigma  -\delta_{\rho\sigma}\gamma_\beta
+\delta_{\sigma\beta} \gamma_\rho).
\label{abj.3}
\end{eqnarray}
Thus after symmetrization associated to the integration over
$q$, the $\xi$ fourth power term is
unique. While lower powers than 4 yield zero under integration 
over $\int d^4\xi$. The terms in eq. (\ref{abj.2}) yield
\begin{eqnarray}&&
(4-D-2)\frac{q^2}{D}
b_\mu \epsilon_{1\rho}\epsilon_{2\sigma}(r+p)_\iota\xi_\mu \xi_\rho
\xi_\sigma \xi_\iota
\nonumber\\&&
+(4-D -2)\frac{q^2}{D}
b_\mu(r-k)_\rho \epsilon_{1\sigma}\epsilon_{2\iota}\xi_\mu \xi_\rho
\xi_\sigma \xi_\iota
\\&&
+(4-D + 2)\frac{q^2}{D}
b_\mu \epsilon_{1\rho}(r)_\sigma\epsilon_{2\iota}
\xi_\mu \xi_\rho \xi_\sigma \xi_\iota.
\label{abj.4}
\end{eqnarray}
Finally
%
%
\begin{eqnarray}&&
\Bigg(
(4-D-2)
 (r+p)_\iota
+(4-D -2)(r-k)_\iota
-(4-D + 2) (r)_\iota
\Bigg)
\nonumber\\&&
\frac{q^2}{D} b_\mu\epsilon_{1\rho}\epsilon_{2\sigma}
\xi_\mu \xi_\rho \xi_\sigma \xi_\iota
\nonumber\\&&
=
\Bigg((4-D-6)r_\iota + (4-D-2)(p-k)_\iota
\Bigg)
\frac{q^2}{D} b_\mu\epsilon_{1\rho}\epsilon_{2\sigma}
\xi_\mu \xi_\rho \xi_\sigma \xi_\iota.
\label{abj.5}
\end{eqnarray}
We insert the value of $r=yk-xp + yp$ and integrate
$\int_0^1 dx \int_0^x dy$
\begin{eqnarray}&&
\Big((4-D-6)r_\iota + (4-D-2)(p-k)_\iota
\Big)
\frac{q^2}{D} b_\mu\epsilon_{1\rho}\epsilon_{2\sigma}
\xi_\mu \xi_\rho \xi_\sigma \xi_\iota
\nonumber\\&&
=\Big(\frac{1}{6}(D+2) + \frac{1}{2}(2-D)
\Big)(p-k)_\iota
\frac{q^2}{D} b_\mu\epsilon_{1\rho}\epsilon_{2\sigma}
\xi_\mu \xi_\rho \xi_\sigma \xi_\iota
\nonumber\\&&
= \frac{1}{3} (4-D) (p-k)_\iota
\frac{q^2}{D} b_\mu\epsilon_{1\rho}\epsilon_{2\sigma}
\xi_\mu \xi_\rho \xi_\sigma \xi_\iota
\label{abj.6}
\end{eqnarray}
which agrees with the derivation of the same expression
in eq. (71) of Ref. \cite{Ferrari:2014jqa}, after integration over
$\int d^4 \xi$.
\par
The remaining terms in eq. (\ref{abj.1}) are finite in the limit $D=4$
therefore we can use the standard algebra with $\gamma_\chi = \gamma_5$.

\section{Example of Ward Identity: ABJ Anomaly}
\label{App:WARD}
We use the Ward identity (\ref{ward.2b}) for the evaluation of the
ABJ anomaly. We start with the amplitude for the triangular graph
\begin{eqnarray}&&
\evr{-}(p+k)_\alpha
\int \frac{d^Dq}{(2\pi)^D}
\frac{Tr\Big(\not\! q\gamma_\rho\cancel{(q-p)}\gamma_\alpha\gamma_\chi
\cancel{(q+k)}\gamma_\sigma\Big)}
{q^2(q-p)^2(q+k)^2}
\nonumber\\&&
\evr{-}(p+k)_\alpha
\int \frac{d^Dq}{(2\pi)^D}
\frac{Tr\Big(\not\! q\gamma_\sigma\cancel{(q-k)}\gamma_\alpha\gamma_\chi
\cancel{(q+p)}\gamma_\rho\Big)}
{q^2(q+p)^2(q-k)^2}
\label{wardabj.1}
\end{eqnarray}
The amplitudes in eq. (\ref{wardabj.1}) transform into each other  under the 
exchange
$\rho,p \leftrightarrow \sigma,k$. Thus we consider only one, the first.
\par 
By using the Ward identity (\ref{ward.2b}) with the momenta
\begin{eqnarray}
-(p+k)_\alpha  \to (q-p)_\alpha - (q+k)_\alpha
\label{wardabj.2}
\end{eqnarray}
we get
\footnote{In fact we should use the procedure outlined in eq. (\ref{intro.001b.2})
and drop the $\gamma_\chi$, since we consider the generic dimension
$D$. We stick to the improper use of $\gamma_\chi$, in order to keep
the notations closer to the common use. The correct notation for $D\sim 4$ would be, for instance,
$Tr\Big(\cancel{(q+k)}\gamma_\sigma\not\! q\gamma_\rho\cancel{(q-p)}\gamma_\alpha\gamma_\chi
\Big)$  $\to$ $\xi(q+k)~\xi_\sigma ~\xi q ~\xi_\rho ~Tr(\cancel{(q-p)}\gamma_\alpha)
+\dots~({\rm permutations}).$
}
\begin{eqnarray}&&
\int \frac{d^Dq}{(2\pi)^D}
{[(q-p)_\alpha-(q+k)_\alpha]}
\frac{Tr\Big(\not\! q\gamma_\rho\cancel{(q-p)}\gamma_\alpha\gamma_\chi
\cancel{(q+k)}\gamma_\sigma\Big)}
{q^2(q-p)^2(q+k)^2}
\nonumber\\&&
=
\int \frac{d^Dq}{(2\pi)^D}
{(q-p)_\alpha}
\frac{Tr\Big(\not\! q\gamma_\rho\cancel{(q-p)}\gamma_\alpha\gamma_\chi
\cancel{(q+k)}\gamma_\sigma\Big)}
{q^2(q-p)^2(q+k)^2}
\nonumber\\&&
+\int \frac{d^Dq}{(2\pi)^D}
{(q+k)_\alpha}
\frac{Tr\Big(\not\! q\gamma_\rho\cancel{(q-p)}\gamma_\chi\gamma_\alpha
\cancel{(q+k)}\gamma_\sigma\Big)}
{q^2(q-p)^2(q+k)^2}
\nonumber\\&&
-\int \frac{d^Dq}{(2\pi)^D}
{(q+k)_\alpha}
\frac{Tr\Big(\not\! q\gamma_\rho\cancel{(q-p)}\{\gamma_\alpha,\gamma_\chi\}
\cancel{(q+k)}\gamma_\sigma\Big)}
{q^2(q-p)^2(q+k)^2}
\label{wardabj.2p}
\end{eqnarray}
By using elementary algebra we get
\begin{eqnarray}&&
=
\int \frac{d^Dq}{(2\pi)^D}
\frac{Tr\Big(\not\! q\gamma_\rho\gamma_\chi
\cancel{(q+k)}\gamma_\sigma\Big)}
{q^2(q+k)^2}
+\int \frac{d^Dq}{(2\pi)^D}
\frac{Tr\Big(\not\! q\gamma_\rho\cancel{(q-p)}\gamma_\chi\gamma_\sigma\Big)}
{q^2(q-p)^2}
\nonumber\\&&
-\int \frac{d^Dq}{(2\pi)^D}
{(q+k)_\alpha}
\frac{Tr\Big(\not\! q\gamma_\rho\cancel{(q-p)}\{\gamma_\alpha,\gamma_\chi\}
\cancel{(q+k)}\gamma_\sigma\Big)}
{q^2(q-p)^2(q+k)^2}
\label{wardabj.3}
\end{eqnarray}
The first two integrals give no contribution to the anomaly
since they depend only on $p$ or on $k$ and not on both.. 
\par
The third integral is the object of our evaluation. The anticommutator
$\{\gamma_\alpha,\gamma_\chi\}$
brings  a factor $D-4$ in the limit. The integral in $q$ is superficially
(logarithmically)) divergent since the $q^4$ term in the numerator
is zero by symmetry. Thus we need only the \underline{pole} part of the
integral over $q$. Finite parts are cancelled by the $D-4$ factor. Ward identity
is very powerful: it tells us what should be evaluated. A small trick 
makes the computation simpler. The use of Feynman parameters and the
change of variable
\begin{eqnarray}
q\to q+r
\label{wardabj.3.1}
\end{eqnarray}
makes the integration on $q$ easier
\begin{eqnarray}&&
-\int \frac{d^Dq}{(2\pi)^D}
{(q+k)_\alpha}
\frac{Tr\Big(\not\! q\gamma_\rho\cancel{(q-p)}\{\gamma_\alpha,\gamma_\chi\}
\cancel{(q+k)}\gamma_\sigma\Big)}
{q^2(q-p)^2(q+k)^2}
\nonumber\\&&
= 
- \int_0^1dx\int_0^{1-x}
\int \frac{d^Dq}{(2\pi)^D}
{(q+\bar k)_\alpha}
\nonumber\\&&
\frac{Tr\Big(\cancel{( q+r)}\gamma_\rho\cancel{(q-\bar p)}\{\gamma_\alpha,\gamma_\chi\}
\cancel{(q+\bar k)}\gamma_\sigma\Big)}
{(q^2-\Delta)^3}
\label{wardabj.3.2}
\end{eqnarray}
where
\begin{eqnarray}&&
\bar k \equiv r+k
\nonumber\\&&
\bar p \equiv p - r .
\label{wardabj.3.3}
\end{eqnarray}
We evaluate the
trace
\begin{eqnarray}&&
{(q+\bar k)_\alpha}
Tr\Big(\cancel{( q+r)}\gamma_\rho\cancel{(q-\bar p)}\{\gamma_\alpha,\gamma_\chi\}
\cancel{(q+ \bar k)}\gamma_\sigma\Big)
\nonumber\\&&
=Tr\Big(\gamma_\chi\Big\{\cancel{(q+\bar k)},\cancel{(q+\bar k)}\gamma_\sigma
 \cancel{( q+r)} \gamma_\rho \cancel{(q-\bar p)}\Big\} \Big).
\label{wardabj.4}
\end{eqnarray}
In the anticommutator we keep only the terms quadratic in $q$ in order to
have a non-zero pole in $D-4$. 
\par
First we evaluate the terms linear in $r$ of eq. (\ref{wardabj.4})
\begin{eqnarray}&&
Tr\Big(\gamma_\chi\Big\{\cancel{(q+\bar k)},\cancel{(q+\bar k)}\gamma_\sigma
 \cancel{r} \gamma_\rho \cancel{(q-\bar p)}\Big\} \Big)
\nonumber\\&&
=
2Tr\Big(\gamma_\chi(q+\bar k)^2\gamma_\sigma
 \cancel{r} \gamma_\rho \cancel{(q-\bar p)} \Big)
\comment{
\nonumber\\&&
\evr{111111111111111111111}}
\nonumber\\&&
-2Tr\Big(\gamma_\chi\cancel{(q+\bar k)}(q+\bar k)_\sigma
\cancel{r} \gamma_\rho \cancel{(q-\bar p)}\Big\} \Big)
\comment{
\nonumber\\&&
\evr{2222222222222222222222}}
\nonumber\\&&
+2Tr\Big(\gamma_\chi\cancel{(q+\bar k)}\gamma_\sigma
(q+\bar k)r \gamma_\rho \cancel{(q-\bar p)}\Big\} \Big)
\comment{
\nonumber\\&&
\evr{3333333333333333333333}}
\nonumber\\&&
-2Tr\Big(\gamma_\chi\cancel{(q+\bar k)}\gamma_\sigma \cancel{r}
(q+\bar k)_\rho \cancel{(q-\bar p)}\Big\} \Big)
\comment{
\nonumber\\&&
\evr{444444444444444444444444444444444444}}
\nonumber\\&&
+2Tr\Big(\gamma_\chi\cancel{(q+\bar k)}\gamma_\sigma \cancel{r}
\gamma_\rho (q+\bar k)(q-\bar p)\Big\} \Big)
\nonumber\\&&
\label{wardabj.4.1}
\end{eqnarray}
%
The quadratic part in $q$ of eq. (\ref{wardabj.4.1}) is
\begin{eqnarray}&&
Tr\Big(\gamma_\chi\Big\{\cancel{(q+\bar k)},\cancel{(q+\bar k)}\gamma_\sigma
 \cancel{r} \gamma_\rho \cancel{(q-\bar p)}\Big\} \Big)
\nonumber\\&&
=
\frac{q^2}{D}   2Tr\Big(\gamma_\chi\gamma_\sigma
 \cancel{r} \gamma_\rho  2 \cancel{\bar k}\Big)
- 2 q^2Tr\Big(\gamma_\chi\gamma_\sigma
 \cancel{r} \gamma_\rho \cancel{\bar p} \Big)
\comment{
\nonumber\\&&
\evr{11111111111111111111111111111111111}}
\nonumber\\&&
-2 \frac{q^2}{D} Tr\Big(\gamma_\chi\cancel{\bar k}\cancel{r} \gamma_\rho
\gamma_\sigma\Big)
+2  \frac{q^2}{D}Tr\Big(\gamma_\chi\gamma_\sigma\cancel{r} \gamma_\rho
\cancel{\bar p} \Big)
\comment{
\nonumber\\&&
\evr{222222222222222222222222222222222}}
\nonumber\\&&
+2\frac{q^2}{D}Tr\Big(\gamma_\chi\cancel{\bar k}\gamma_\sigma\gamma_\rho
\not\! r  \Big)
-2\frac{q^2}{D}Tr\Big(\gamma_\chi\not\! r\gamma_\sigma\gamma_\rho\cancel{\bar p} \Big)
\comment{
\nonumber\\&&
\evr{3333333333333333333333333333333333333}}
\nonumber\\&&
-2\frac{q^2}{D}Tr\Big(\gamma_\chi\cancel{\bar k}\gamma_\sigma\not\! r 
\gamma_\rho \Big)
+ 2\frac{q^2}{D}Tr\Big(\gamma_\chi\gamma_\rho \gamma_\sigma\not\! r \cancel{\bar p}
 \Big)
\comment{
\nonumber\\&&
\evr{4444444444444444444444444444444444444444444}}
\nonumber\\&&
+2 q^2Tr\Big(\gamma_\chi\cancel{\bar k}\gamma_\sigma\not\! r \gamma_\rho\Big)
+2\frac{q^2}{D}Tr\Big(\gamma_\chi\cancel{(\bar k -\bar p)}
\gamma_\sigma\not\! r \gamma_\rho \Big)
\label{wardabj.4.2}
\end{eqnarray}
Eq. (\ref{wardabj.4.2}) yields
\begin{eqnarray}&&
Tr\Big(\gamma_\chi\gamma_\sigma
 \cancel{r} \gamma_\rho   \cancel{\bar k}\Big)\Big[
\frac{4}{D} +\frac{2}{D}+\frac{2}{D}+\frac{2}{D} -2-\frac{2}{D}
\Big]
\nonumber\\&&
+Tr\Big(\gamma_\chi\gamma_\sigma
 \cancel{r} \gamma_\rho   \cancel{\bar p}\Big)\Big[
 -2+\frac{2}{D}+\frac{2}{D}+\frac{2}{D}+\frac{2}{D}
\Big]
\nonumber\\&&
=Tr\Big(\gamma_\chi\gamma_\sigma
 \cancel{r} \gamma_\rho   \cancel{\bar k}\Big)
\frac{2}{D}(4-D)
\nonumber\\&&
+Tr\Big(\gamma_\chi\gamma_\sigma
 \cancel{r} \gamma_\rho   \cancel{\bar p}\Big)
\frac{2}{D}(4-D)
\nonumber\\&&
=\frac{2}{D}(4-D)Tr\Big(\gamma_\chi\gamma_\sigma
 \cancel{r} \gamma_\rho   \cancel{(\bar k + \bar p)}\Big)
\nonumber\\&&
=\frac{2}{D}(4-D)Tr\Big(\gamma_\chi\gamma_\sigma
 \cancel{r} \gamma_\rho   \cancel{( k +  p)}\Big)
\label{wardabj.4.3}
\end{eqnarray}
After integration over Feynman parameters we have
\begin{eqnarray}&&
2\int_0^1dx \int_0^{1-x} dy =1
\nonumber\\&&
2\int_0^1dx \int_0^{1-x} dy r =2\int_0^1dx \int_0^{1-x} dy (y k- xp)
= \frac{1}{3}(k-p)
\label{wardabj.4.4}
\end{eqnarray}
Then the mean value of the quantity in eq. (\ref{wardabj.4.3}) is
\begin{eqnarray}&&
2\int_0^1dx \int_0^{1-x} dy\frac{2}{D}(4-D)Tr\Big(\gamma_\chi\gamma_\sigma
 \cancel{r} \gamma_\rho   \cancel{( k +  p)}\Big)
\nonumber\\&&
 =
\frac{1}{3}\frac{2}{D}(4-D)Tr\Big(\gamma_\chi\gamma_\sigma
 \cancel{(k-p)} \gamma_\rho   \cancel{( k +  p)}\Big)
\nonumber\\&&
= \frac{4}{3}\frac{4-D}{D}Tr\Big(\gamma_\chi\gamma_\sigma
\not \!k \gamma_\rho   \not\!p\Big)
\label{wardabj.4.5}
\end{eqnarray}
Now we have to remember of a factor $2$ coming from the second
graph in eq. (\ref{wardabj.1}) and the integration over 
$q$ i.e. $\int d^Dq (2\pi)^{-D}\sim 2i (4-D)^{-1 } (4\pi)^{-2}$.
Thus we get
\begin{eqnarray}&&
 \frac{4}{3}\frac{4-D}{D}Tr\Big(\gamma_\chi\gamma_\sigma
\not \!k \gamma_\rho   \not\!p\Big)4i \frac{1}{4-D}\frac{1}{(4\pi)^2}
\nonumber\\&&
=
\frac{i}{(4\pi)^2}\frac{4}{3}Tr\Big(\gamma_\chi\gamma_\sigma
\not \!k \gamma_\rho   \not\!p\Big)
\label{wardabj.4.6}
\end{eqnarray}
%
\par
Now we evaluate the second term of eq. (\ref{wardabj.4})
\begin{eqnarray}&&
=2Tr\gamma_\chi\Big((q+k)^2 \gamma_\sigma
 \not\! q \gamma_\rho \cancel{(q-p)}
\nonumber\\&&
-\cancel{(q+k)}(q+k)_\sigma  \not\! q \gamma_\rho \cancel{(q-p)}
\nonumber\\&&
+\cancel{(q+k)}\gamma_\sigma\gamma_\rho \cancel{(q-p)} (q+k)q
\nonumber\\&&
-\cancel{(q+k)}\gamma_\sigma \not\! q (q+k)_\rho\cancel{(q-p)}
\nonumber\\&&
+\cancel{(q+k)}\gamma_\sigma \not\! q \gamma_\rho (q+k)(q-p) \Big)
\label{wardabj.5}
\end{eqnarray}
We evaluate the terms in the RHS of eq. (\ref{wardabj.5}) one by one
\begin{eqnarray}&&
2Tr\gamma_\chi\Big((q+k)^2 \gamma_\sigma
 \not\! q \gamma_\rho \cancel{(q-p)} \Big)
= -2Tr\gamma_\chi\Big(2qk \gamma_\sigma
 \not\! q \gamma_\rho \cancel{p} \Big)
\nonumber\\&&
= -2\frac{2q^2}{D}Tr\gamma_\chi\Big( \gamma_\sigma
 \not\! k \gamma_\rho \cancel{p} \Big)
\label{wardabj.6}
\end{eqnarray}
\begin{eqnarray}&&
2Tr\gamma_\chi\Big(
-\cancel{(q+k)}(q+k)_\sigma  \not\! q \gamma_\rho \cancel{(q-p)} \Big)
\nonumber\\&&
=2Tr\gamma_\chi\Big(\cancel{(q+k)}(q+k)_\sigma  \not\! q \gamma_\rho \cancel{p} \Big)
=2Tr\gamma_\chi\Big(\frac{q^2}{D}\cancel{k}\gamma_\sigma   \gamma_\rho \cancel{p} \Big)
\label{wardabj.7}
\end{eqnarray}
\begin{eqnarray}&&
2Tr\gamma_\chi\Big(
\cancel{(q+k)}\gamma_\sigma\gamma_\rho \cancel{(q-p)} (q+k)q \Big)
\nonumber\\&&
=
 q^2
2Tr\gamma_\chi\Big(
\cancel{(q+k)}\gamma_\sigma\gamma_\rho \cancel{(q-p)} \Big)
+2Tr\gamma_\chi\Big(
\cancel{(q+k)}\gamma_\sigma\gamma_\rho \cancel{(q-p)} qk \Big)
\nonumber\\&&
=
-2q^2Tr\gamma_\chi\Big(\not\! k\gamma_\sigma\gamma_\rho \not\! p \Big)
-2\frac{q^2}{D}Tr\gamma_\chi\Big(
\not\! k\gamma_\sigma\gamma_\rho \not\!p \Big)
\label{wardabj.8}
\end{eqnarray}
\begin{eqnarray}&&
2Tr\gamma_\chi\Big(
-\cancel{(q+k)}\gamma_\sigma \not\! q (q+k)_\rho\cancel{(q-p)}\Big)
\nonumber\\&&
=
2 \frac{q^2}{D}
Tr\gamma_\chi\Big(
\not\!k\gamma_\sigma \gamma_\rho\not\! p\Big)
\label{wardabj.9}
\end{eqnarray}
\begin{eqnarray}&&
2Tr\gamma_\chi\Big(
\cancel{(q+k)}\gamma_\sigma \not\! q \gamma_\rho (q+k)(q-p) \Big)
\nonumber\\&&
=
-2\frac{q^2}{D}
Tr\gamma_\chi\Big(
 \not\! k\gamma_\sigma \not\! p \gamma_\rho  \Big)
\label{wardabj.10}
\end{eqnarray}
The sum of the five terms yields
\begin{eqnarray}&&
(\ref{wardabj.3})
=
-\int \frac{d^Dq}{(2\pi)^D} q^2 (\frac{4}{D}-1)\frac{
2Tr\gamma_\chi\Big(
 \not\! k\gamma_\sigma \gamma_\rho  \not\! p \Big)}
{q^2(q-p)^2(q+k)^2}
\nonumber\\&&
=  \frac{i}{(4\pi)^2}\frac{2}{4-D}(\frac{4}{D}-1)2
Tr\gamma_\chi\Big(
 \not\! k\gamma_\sigma \gamma_\rho  \not\! p \Big)
\nonumber\\&&
= \frac{i}{(4\pi)^2}Tr\gamma_\chi\Big(
 \not\! k\gamma_\sigma \gamma_\rho  \not\! p \Big).
\label{wardabj.13}
\end{eqnarray}
Finally we add the contribution (\ref{wardabj.4.6})
of the $q$ part
of eq. (\ref{wardabj.4}) and get the axial anomaly
\begin{eqnarray}&&
2\frac{i}{(4\pi)^2}Tr\gamma_\chi\Big(
 \not\! k\gamma_\sigma \gamma_\rho  \not\! p \Big)
+\frac{i}{(4\pi)^2} \frac{4}{3}  Tr\gamma_\chi\Big(\gamma_\sigma
 \not\! k \gamma_\rho  \not\! p \Big) 
\nonumber\\&&
=  \frac{2}{3}\frac{i}{(4\pi)^2}Tr\gamma_\chi\Big(
 \not\! k\gamma_\sigma \gamma_\rho  \not\! p \Big),
\label{wardabj.14}
\end{eqnarray}
which is in agreement with the standard result.
\par
With this method the derivation of the anomaly is really
very simple: by matching the pole's and the zero's
one avoids a lot of of unnecessary algebra. The method
is supported by the fact that $\{\gamma_\alpha,\gamma_\chi\}$
has a zero for $D\to 4$ and finally by the existence 
of an integral representation of the gamma's algebra, i.e.
eqs. (\ref{intro.001a}) and  (\ref{intro.001b}).


\begin{thebibliography}{99}
\small




\bibitem{Ferrari:2014jqa} 
  R.~Ferrari,
  ``Managing $\gamma_5$ in Dimensional Regularization and ABJ Anomaly,''
  arXiv:1403.4212 [hep-th].

\bibitem{berezin}
F.~A.~Berezin, ``The Method of Second Quantization, `` 
Academic Press, New York, 1966.

\bibitem{E.R.Caianiello:1952ww} 
  S.~Fubini and E.~R.~Caianiello,
  ``On the Algorithm of Dirac spurs,''
  Nuovo Cim.\  {\bf 9}, 1218 (1952).


\bibitem{Jaffe:1987bf} 
  A.~M.~Jaffe, A.~Lesniewski and J.~Weitsman,
  ``Pfaffians On Hilbert Space,''
  J. Funct. Anal.\ {\bf 83}, 348 (1989). 
  HUTMP-87/B207.



\bibitem{Adler:1969gk} 
  S.~L.~Adler,
  ``Axial vector vertex in spinor electrodynamics,''
  Phys.\ Rev.\  {\bf 177}, 2426 (1969).


\bibitem{Bell:1969ts}
  J.~S.~Bell and R.~Jackiw,
  ``A PCAC puzzle: pi0 $\to$ gamma gamma in the sigma model,''
  Nuovo Cim.\ A {\bf 60} (1969) 47.








\bibitem{Ferrari:2005ii}
  R.~Ferrari,
  ``Endowing the nonlinear sigma model with a flat connection structure: A  way
  to renormalization,''
  JHEP {\bf 0508}, 048 (2005)
  [arXiv:hep-th/0504023].

\bibitem{Ferrari:2005va}
  R.~Ferrari and A.~Quadri,
  ``A weak power-counting theorem for the renormalization of the non-linear
  sigma model in four dimensions,''
  Int.\ J.\ Theor.\ Phys.\  {\bf 45}, 2497 (2006)
  [arXiv:hep-th/0506220].

\bibitem{Bettinelli:2007zn}
  D.~Bettinelli, R.~Ferrari and A.~Quadri,
  ``Further comments on the renormalization of the nonlinear sigma model,''
  Int.\ J.\ Mod.\ Phys.\  A {\bf 23}, 211 (2008)
  [arXiv:hep-th/0701197].

\bibitem{Bettinelli:2007kc}
  D.~Bettinelli, R.~Ferrari and A.~Quadri,
  ``Path-integral over non-linearly realized groups and hierarchy solutions,''
  JHEP {\bf 0703} (2007) 065
  [arXiv:hep-th/0701212].



\bibitem{Ferrari:2015mha} 
  R.~Ferrari,
  ``Managing $\gamma_5$ in Dimensional Regularization II: the Trace with more $\gamma_{5}$$\prime s$,''
  Int.\ J.\ Theor.\ Phys.\  {\bf 56}, no. 3, 691 (2017)
  doi:10.1007/s10773-016-3211-8
  [arXiv:1503.07410 [hep-th]].



\bibitem{Bardeen:1969md} 
  W.~A.~Bardeen,
  ``Anomalous Ward identities in spinor field theories,''
  Phys.\ Rev.\  {\bf 184}, 1848 (1969).
  doi:10.1103/PhysRev.184.1848


\bibitem{bertlmann}
 R.~A:~ Bertlmann, ``Anomalies in Quantum Field Theories, ``
   Clarendon Press, Oxford 1996.



\bibitem{Ferrari:2015xba} 
  R.~Ferrari and M.~Raciti,
  ``On effective Chern-Simons Term induced by a Local CPT-Violating Coupling using $\gamma_5$ in Dimensional Regularization,''
  arXiv:1510.04666 [hep-th].


\bibitem{Jackiw:1999yp} 
  R.~Jackiw and V.~A.~Kostelecky,
  ``Radiatively induced Lorentz and CPT violation in electrodynamics,''
  Phys.\ Rev.\ Lett.\  {\bf 82}, 3572 (1999)
  [hep-ph/9901358].

\bibitem{Jackiw:1999qq} 
  R.~Jackiw,
  ``When radiative corrections are finite but undetermined,''
  Int.\ J.\ Mod.\ Phys.\ B {\bf 14}, 2011 (2000)
  [hep-th/9903044].

\bibitem{El-Menoufi:2015dra} 
  B.~K.~El-Menoufi and G.~A.~White,
  arXiv:1505.01754 [hep-th].

\end{thebibliography}
\end{document}